%% file: main-arxiv.tex
\documentclass{llncs}
\usepackage{paralist}

\usepackage{cite}

\usepackage{footmisc}

\makeatletter

\renewcommand\section{\@startsection{section}{1}{\z@}%
{-10\p@ \@plus -4\p@ \@minus -4\p@}
{3\p@ \@plus 2\p@ \@minus 2\p@}
{\normalfont\large\bfseries\boldmath
\rightskip=\z@ \@plus 8em\pretolerance=10000 }}
\renewcommand\subsection{\@startsection{subsection}{2}{\z@}%
{-6\p@ \@plus -4\p@ \@minus -4\p@}
{2\p@ \@plus 4\p@ \@minus 4\p@}
{\normalfont\normalsize\bfseries\boldmath
\rightskip=\z@ \@plus 8em\pretolerance=10000 }}
\renewcommand\subsubsection{\@startsection{subsubsection}{3}{\z@}%
{-3\p@ \@plus -4\p@ \@minus -4\p@}
{-0.5em \@plus -0.22em \@minus -0.1em}
{\normalfont\normalsize\bfseries\boldmath}}
\renewcommand\paragraph{\@startsection{paragraph}{4}{\z@}%
{-4\p@ \@plus -4\p@ \@minus -4\p@}
{-0.5em \@plus -0.22em \@minus -0.1em}
{\normalfont\normalsize\bfseries\boldmath}}

\usepackage[T1]{fontenc}
\usepackage{graphicx}
\input{commands}
\begin{document}
%
\title{Transforming RDF-star to Property Graphs: \\ A Preliminary Analysis of Transformation Approaches -- extended version}
\titlerunning{Transforming RDF-star to Property Graphs}
%
\author{Ghadeer Abuoda\inst{1} \and Daniele Dell'Aglio\inst{1} \and Arthur Keen\inst{2} \and
Katja Hose\inst{1}}
%
\authorrunning{G. Abuoda et al.}
%
\institute{Department of Computer Science, Aalborg University, Aalborg, Denmark\\
\email{\{gsmas,dade,khose\}@cs.aau.dk} \and 
ArangoDB, San Francisco, United States\\
\email{arthur@arangodb.com}}
\maketitle              
\begin{abstract}
\input{abstract}
\end{abstract}

\setcounter{footnote}{0} 

\section{Introduction}
\label{sec:introduction}
\input{introduction-new}

\section{Preliminaries}
\label{sec:preliminaries}
\input{preliminaries}

\section{Related Work}
\label{sec:relatedWork}
\input{relatedWork}
\section{Transformation Approaches from RDF to Property Graphs}
\label{sec:transformation}
\input{transformation-new}

\section{Test Cases}
\label{sec:cases}
\input{casesws}
\section{Analysis and Discussion}
\label{sec:experiments}
\input{discussionws}

\subsection{Discussion}
\input{approachws}

\section{Conclusion}
\label{sec:conclusion}
\input{conclusion}
\subsubsection*{Acknowledgements} 
This research was partially funded by the Danish Council for Independent Research (DFF) under grant agreement no. DFF-8048-00051B and the Poul Due Jensen Foundation.

%
%
%
 \bibliographystyle{splncs04}
 \bibliography{bibliography}

\clearpage
\appendix

 \section{Test Cases}
 \label{App:cases}
\input{cases}
\section{Analysis Output}
\label{App:results}
\input{experiments}
%

\end{document}

%% file: commands.tex
\usepackage{soul,color}
\usepackage{textcomp}
\usepackage{xcolor}
\usepackage{listings}
\usepackage{verbatim}
\usepackage{booktabs}
\usepackage{multirow}
\newcommand{\stitle}[1]{\vspace{1ex}\noindent{\bf #1}}

\newcommand{\ntitle}[1]{\vspace{1ex}\noindent{#1}}

\usepackage{listingsutf8}
\usepackage{float}
\newsavebox{\algleft}
\newsavebox{\algright}

\usepackage[labelfont=bf]{subcaption}
\usepackage{lipsum}

\usepackage{hyperref}

\usepackage[T1]{fontenc}
\usepackage{upquote}
\usepackage{ltablex}
\usepackage{longtable}
\usepackage{tabularx}
\usepackage{ltxtable} 
\usepackage{xtab,afterpage}
\usepackage{lipsum} 
\usepackage{multicol}

\usepackage{pdflscape}
\usepackage{afterpage}
\usepackage{algorithmic}

\usepackage[pass]{geometry}
\usepackage{amsmath}
\usepackage{amsfonts}
\usepackage{amssymb}

\usepackage{xcolor} 
\usepackage[ruled]{algorithm2e}
\usepackage{tikz}
\usetikzlibrary{fit,calc}

\colorlet{pink}{red!40}

\def\HiLi{\leavevmode\rlap{\hbox to 0.8\linewidth{\color{blue!14}\leaders\hrule height 1\baselineskip depth .5ex\hfill}}}

\SetCommentSty{mycommfont}

\newcommand{\CASE}[1]{\STATE \textbf{case} #1\textbf{:} \begin{ALC@g}}
\newcommand{\ENDCASE}{\end{ALC@g}}

\DeclareCaptionFont{white}{\color{white}} 
\DeclareCaptionFormat{listing}{%
  \hspace*{-0.4pt}\colorbox{white}{\parbox{\dimexpr\textwidth-2\fboxsep+.8pt\relax}{#1#2#3}}} 
\usepackage{ragged2e}
\newcolumntype{P}[1]{>{\RaggedRight\arraybackslash}p{#1}}

\definecolor{mygray}{RGB}{239, 239, 239}
\definecolor{mygreen}{rgb}{0,0.6,0}
\lstdefinestyle{st:rdf}{
backgroundcolor =\color{mygray},
language=SQL,
breaklines=true,
showspaces=false,
otherkeywords={GRAPH,rdf:type, rdf:, rdfs:, xsd:, ex:, @prefix,info},
keywordstyle=\color{blue},
basicstyle=\ttfamily\scriptsize,
commentstyle=\color{mygreen},
comment = [f]{\#},
}

\usepackage{parallel,enumitem}

\newcommand{\setalglineno}[1]{%
  \setcounter{ALC@line}{\numexpr#1-1}}

%% file: abstract.tex

RDF and property graph models have many similarities, such as using basic graph concepts like nodes and edges. However, such models differ in their modeling approach, expressivity, serialization, and the nature of applications. 
RDF is the de-facto standard model for knowledge graphs on the Semantic Web and supported by a rich ecosystem for inference and processing. 
The property graph model, in contrast, provides advantages in scalable graph analytical tasks, such as graph matching, path analysis, and graph traversal. 
RDF-star extends RDF and allows capturing metadata as a first-class citizen.
To tap on the advantages of alternative models, the literature proposes different ways of transforming knowledge graphs between property graphs and RDF. 
However, most of these approaches cannot provide complete transformations for RDF-star graphs. 
Hence, this paper provides a step towards transforming RDF-star graphs into property graphs. 
In particular, we identify different cases to evaluate transformation approaches from RDF-star to property graphs. 
Specifically, we categorize two classes of transformation approaches and analyze them based on the test cases.
The obtained insights will form the foundation for building complete transformation approaches in the future. 

%% file: introduction-new.tex

The most popular models for representing knowledge graphs are: RDF\footnote{\scriptsize RDF 1.1 Primer: \url{https://www.w3.org/TR/rdf11-primer/} \label{fn:rdf}} (Resource Description Framework) and property graphs~\cite{rodriguez2010constructions} (PG). 
While RDF represents knowledge graphs as a set of subject-predicate-object triples, property graphs assign key-value style properties to nodes and edges.
Recently, RDF-star~\cite{hartig2014foundations} has been proposed as an extension of RDF to enable enriching RDF triples with metadata information by embedding triples in subjects or objects of other triples, which allows providing statements about statements and somewhat resembles adding properties to edges in property graphs. 
RDF-star is supported by a rich ecosystem of data management systems and standards, most notably systems such as Stardog, 
OpenLink’s Virtuoso, 
Ontotext GraphDB, 
AllegroGraph, 
Apache Jena, 
and more recently also Oxigraph,  
but also query standards, such as SPARQL\footnote{\scriptsize SPARQL 1.1 Query Language: \url{https://www.w3.org/TR/sparql11-query/}} and its extension SPARQL-star\footnote{\scriptsize SPARQL-star Query Language: \url{https://w3c.github.io/rdf-star/cg-spec/editors_draft.html\#sparql-star}} as well as RDF Schema, which allows describing classes of RDF resources and properties\footnote{\scriptsize RDF Schema 1.1: \url{https://www.w3.org/TR/rdf-schema/}}. In contrast, many graph database systems, such as Neo4j, 
TigerGraph, 
JanusGraph,  
RedisGraph, 
and SAP HANA are based on different variations of the property graph model~\cite{tomaszuk2016rdf} and different query languages~\cite{cypher,deutsch2020aggregation}. 
Unfortunately, RDF-star graphs and property graphs are not entirely compatible with one another. Although they both describe data through graphs, their underlying models and semantics are different, leading to many data interoperability issues~\cite{angles2019rdf,lassila2021graph}. 
Metadata or edge properties in RDF-star can be modeled as separate nodes or RDF-star triples.
In contrast, edge properties can only be represented as literal key-value pairs in property graphs. 
In general, it is challenging to transform an RDF-star graph fully into a property graph because of the rich expressiveness of the former. 
The heterogeneity between the two models and their frameworks makes it necessary to study their interoperability, i.e., the ability to map one model to another for data exchange and sharing~\cite{angles2019rdf}. 

The mapping between the two models is crucial for data exchange, data integration 
as well as reusability of systems and tools between the frameworks. 
RDF-star, specifically the RDF model, is recognized as a web-native model that supports data exchange and sharing across different sources because of its formal semantics and the universal uniqueness of resources using IRIs. 
RDF is a common and flexible model for knowledge representation, and that is exemplified by knowledge graphs that cover a broad set of domains, such as DBpedia~\cite{lehmann2015dbpedia}, YAGO~\cite{suchanek2007yago}, and Wikidata~\cite{vrandevcic2012wikidata}. 
On the contrary, even with the wide adoption of property graph engines, property graphs lack many essential features, such as a schema language, a standard query language, standard data serialization formats, etc. 
Achieving interoperability and reliable transformations between the two frameworks will finally enable us to exploit the benefits of both models.

\begin{figure*}[t]
\makebox[\linewidth][c]{%
\begin{subfigure}[b]{.4\textwidth}
\centering
\includegraphics[width=.90\textwidth]{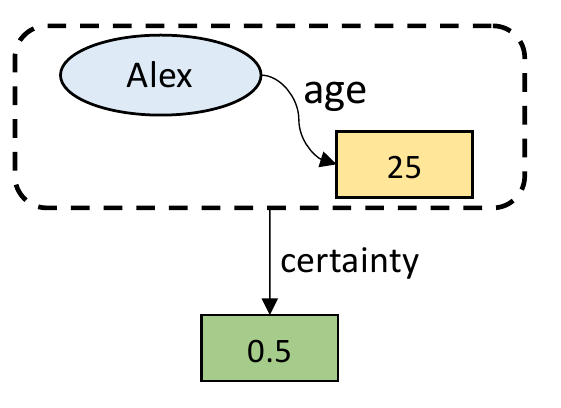}
\caption{\textbf{RDF-star Graph}}
\end{subfigure}%
\begin{subfigure}[b]{.37\textwidth}
\centering
\includegraphics[width=.90\textwidth]{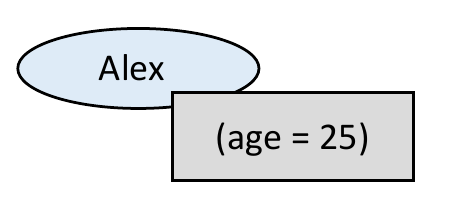}
\caption{\textbf{A property graph}}
\end{subfigure}
}
\caption{Graphical Representation of Listing~\ref{lst:motivation} as (a) RDF-star and (b) Property Graph}
\label{fig:motivationExample}
\end{figure*}

The transformation of property graphs to RDF-star has been explored recently~\cite{khayatbashi2022converting}, and basic transformation rules for property graphs to RDF-star were proposed~\cite{hartig2014foundations,angles2020mapping}. 
However, the latter does not cover all RDF-star constructs and allows for multiple alternatives. 

\begin{lstlisting}[
style={st:rdf},
caption={An example RDF-star graph in Turtle star format},
label={lst:motivation}
]
@prefix ex: <http://example.org/> .
<<ex:Alex ex:age 25>>  ex:certainty 0.5 . 
\end{lstlisting}

Consider, for instance, the example illustrated in Figure~\ref{fig:motivationExample}(a).
If we start with the triple (\texttt{ex:Alex}, \texttt{ex:age}, \texttt{25}), then we could represent the RDF element (\url{Alex}) as a node in a property graph, as shown in Figure~\ref{fig:motivationExample}(b). 
This node would then have a property \url{(age, 25)} and the RDF triple would be represented by a single node in a property graph. 
However, if we have a single node without an edge, we cannot represent the metadata about the original RDF triple (ex:certainty 0.5) in the property graph. 
Studying such cases, this paper makes the following contributions:
\begin{itemize}
    \item We identify two alternative approaches of transformations: RDF-topology-preserving and Property-Graph transformation.
    \item We define a set of test cases capturing the diverse RDF-star constructs that have to be considered when transforming RDF-star to property graphs.
    \item Using the test cases, we systematically evaluate alternative mapping approaches and identify their shortcomings. 
\end{itemize}
This paper is structured as follows: while Section~\ref{sec:preliminaries} introduces preliminaries, Section~\ref{sec:relatedWork} discusses related work. 
Section~\ref{sec:transformation} presents alternative transformation approaches. 
Afterwards, Section~\ref{sec:cases} provides details on our test cases, which we use in Section~\ref{sec:experiments} to identify and discuss shortcomings of transformation approaches. 
Section~\ref{sec:conclusion} concludes the work with an outlook for future work. 

%% file: preliminaries.tex

In this section, we formally introduce RDF\footref{fn:rdf}, RDF-star~\cite{hartig2014foundations}, and property graphs~\cite{rodriguez2010constructions}. 

\subsubsection{Resource Description Framework (RDF). }
RDF 
is a W3C standard data model that represents information as a set of statements. Each statement denotes a typed relation between two resources. 

\begin{definition}[\textbf{RDF statement}]
Let $I$, $B$ and $L$ be the disjoint sets of Internationalized Resource Identifiers (IRIs), blank nodes and literals. 
An \emph{RDF statement} is a triple $(s,p,o) \in (I \cup B)\times I \times (I\cup B \cup L)$, and it indicates that $s$ and $o$ (\emph{subject} and \emph{object}, resp.) are in a relation $p$ (\emph{predicate}). 
\end{definition}

\begin{figure}[t]
\makebox[\linewidth][c]{%
\begin{subfigure}[b]{.5\textwidth}
\centering
\includegraphics[width=.93\textwidth]{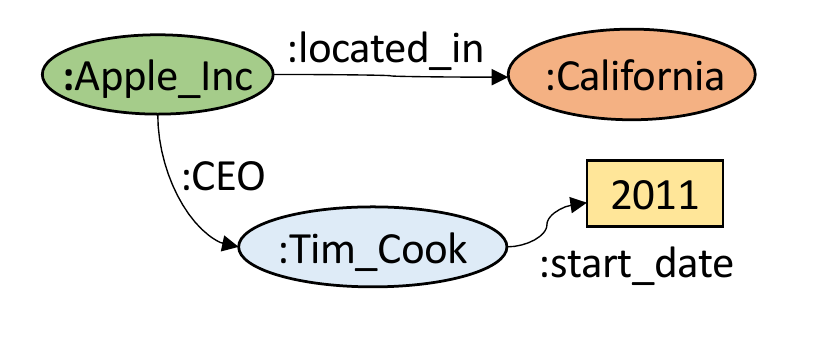}
\caption{\textbf{RDF Graph}}
\end{subfigure}%
\begin{subfigure}[b]{.5\textwidth}
\centering
\includegraphics[width=.93\textwidth]{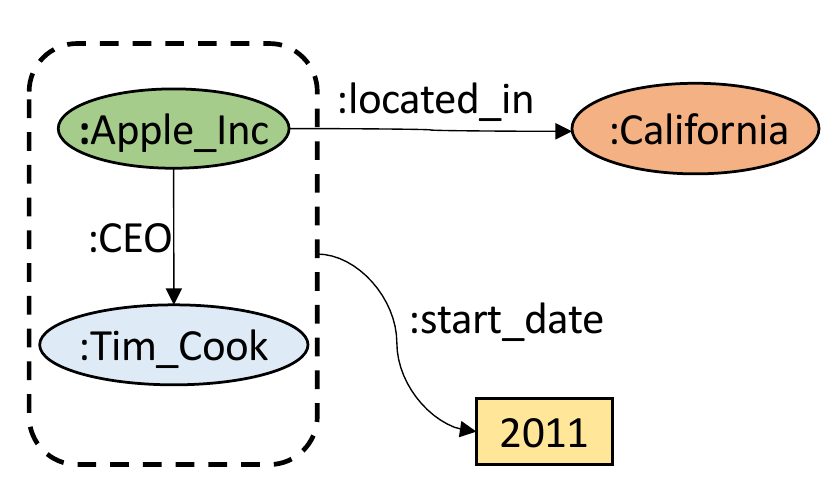}
\caption{\textbf{RDF-star Graph}}
\end{subfigure}
\begin{subfigure}[b]{.5\textwidth}
\centering
\includegraphics[width=.93\textwidth]{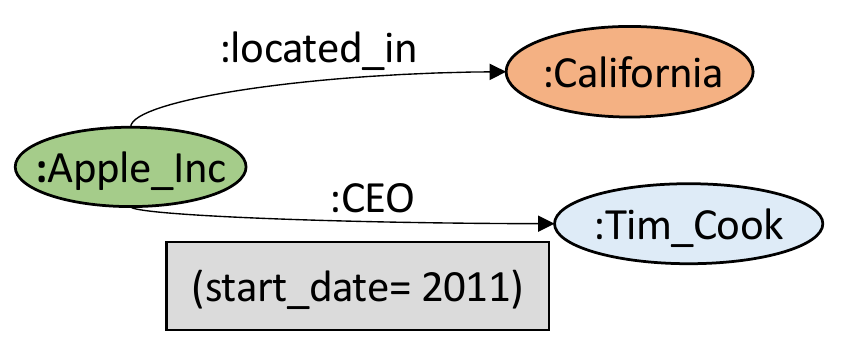}
\caption{\textbf{Property Graph}}
\end{subfigure}%
}
\caption{Example in RDF, RDF-star, and Property Graphs}
\label{fig:rdfpgExample}
\end{figure}

In this paper, we consider two types of RDF statements that we distinguish based on whether the object is an IRI or a literal. \emph{Object property statements} are RDF statements $(s,p,o) \in (I\cup B) \times I \times (I \cup B)$, while \emph{datatype property statements} are RDF statements $(s,p,o) \in (I\cup B) \times I \times L$.
An RDF graph containing three RDF statements is shown in Listing~\ref{lst:rdfTurtle} serialized in Turtle\footnote{\scriptsize RDF Turtle: \url{https://www.w3.org/TR/turtle/}}, and visually in Figure~\ref{fig:rdfpgExample}(a) as a graph. 
The first two statements are object property statements. The first statement describes two resources, \url{ex:Apple\_Inc} and \url{ex:California}, related by the predicate  \url{ex:located\_in}. 
The second statements indicates that \url{ex:Apple\_Inc} has \url{ex:Tim\_Cook} as a \url{ex:CEO}. 
The last statement is a datatype property statement, and it indicates that \url{ex:Tim\_Cook} has the literal "2011" as the value of the \url{ex:start\_date} predicate.

\begin{lstlisting}[
style={st:rdf},
caption={An RDF graph in Turtle format},
label={lst:rdfTurtle}
]
@prefix ex: <http://example.org/> .
ex:Apple_Inc ex:located_in ex:California .
ex:Apple_Inc ex:CEO ex:Tim_Cook .
ex:Tim_Cook ex:start_date 2011 .
\end{lstlisting}

\subsubsection{RDF-star} Looking at the RDF graph in Figure~\ref{fig:rdfpgExample}(a), one can spot some imprecise data modelling choices: stating that Tim Cook started in 2011 is not totally correct, as he started in 2011 his role as CEO of Apple. 
In other words, one should associate the starting date to the statement (\url{ex:Apple\_Inc}, \url{ex:CEO}, \url{ex:Tim\_Cook}), as depicted in Figure~\ref{fig:rdfpgExample}(b). 
There are several ways to implement this idea in RDF, such as RDF reification~\cite{hartig2014foundations}, singleton properties~\cite{nguyen2014don}, and named graphs~\cite{carroll2005named}. However, these mechanisms have significant shortcomings~\cite{metadataEvalution,RDFReconciliation,HernandezGH21,PelgrinGH21}.

\begin{lstlisting}[
style={st:rdf},
caption={An RDF-Star Graph in Turtle-Star Format},
label={lst:starTurtle}
]
@prefix ex: <http://example.org/> .
  ex:Apple_Inc ex:located_in ex:California . 
<<ex:Apple_Inc ex:CEO    ex:Tim_Cook>>  ex:start_date 2011 .
\end{lstlisting}

A solution to overcome such shortcomings was recently proposed by Hartig et al. with RDF-star~\cite{hartig2014foundations,RDFReconciliation}. 
RDF-star extends RDF by letting RDF statements be subjects or objects in other statements. 
Listing~\ref{lst:starTurtle} shows an RDF-star document serialized in Turtle-star~\cite{hartig2014foundations}. 
The first statement is a compliant RDF statement (it appeared also in Listing \ref{lst:rdfTurtle}). The second statement indicates that \url{ex:Apple\_Inc} appointed \url{ex:Tim\_Cook} as \url{ex:CEO} in 2011. 
We formally define an RDF-star statement as follows.

\begin{definition}[\textbf{RDF-star statement}] Let $s \in I \cup B$ , $p \in I$, $o \in I \cup B \cup L$. An RDF-star statement is a triple defined recursively as:
\begin{itemize}
\item Any RDF statement $(s, p, o)$ is an RDF-star statement;
\item  Let $t$ and $\bar{t}$ be RDF-star statements. Then, $(t, p, o)$, $(s, p, t)$ and $(t, p, \bar{t})$ are RDF-star statements, also known as \emph{asserted statement}. $t$ and $\bar{t}$ are called \emph{embedded} or \emph{quoted} statements.
\end{itemize}
\end{definition}

\subsubsection{Property Graphs}
A property graph (PG) is a graph where  nodes and edges can have multiple properties, represented as key-value pairs. Figure~\ref{fig:rdfpgExample}(c) illustrates the graph described in the above section as a property graph.
In this case, the starting date of Tim Cook as the CEO of Apple Inc is reported as a key-value property on the \url{:CEO} edge.
%
PGs have not a unique and standardized model; each PG engine proposes its data model. 
A generic PG model definition is proposed by~\cite{tomaszuk2016rdf}.
\begin{definition}[\textbf{Property Graph}]
Let $L$ be the set of the labels, $PN$ be the set of property names, and $D$ be the set of property values.
A property graph $G$ is an edge-labeled directed multi-graph such that $G = (N, E, edge, lbl, P, \sigma)$, where:
\begin{itemize}
    \item $N$ is a set of nodes, 
    \item $E$ is a set of edges between nodes, such that $N \cap E =\emptyset$
    \item $edge : E \to(N \times N) $ is a total function that associates each edge in $E$ with a pair of nodes in $N$. If $edge(e_{1}) = (n_{1}, n_{2})$, 
    $n_{1}$ is the source node and $n_{2}$ is the target node. 
    \item  $lbl : (N\cup E) \rightarrow \mathcal{P}($L$)$ is a function that associates each edge or node with a set of labels.
    \item $\sigma: (N \cup E) \rightarrow \mathcal{P}(P)$ is a function that associates a node or edge with a non-empty set of properties $P$ defined as a set of key-value pairs ($k,v$) where $k \in PN$ and $v \in D$
  \end{itemize}  
\end{definition}
To ease all approaches' output representation, we map any IRI to a distinct string representing a local name. Given $I$, a set of all IRIs, $localName$ is a function that maps an IRI to a string that represents the local name of an RDF resource\footnote{In Neo4j, the user can configure the local name of RDF terms such as subPropertyOf, subClassOf, Class, etc.}. For example, the local name for the RDF resource (\url{http://example.com/meets}) $localName$("\url{http://example.com/meets}") is "\texttt{meets}". We will use this function in the output representation in Section~\ref{sec:experiments}.

%% file: relatedWork.tex

We can distinguish between related work on converting between (i) RDF and PG and (ii) RDF-star and PG.

\paragraph{RDF and PG. }
%
Angles et al.~\cite{angles2020mapping} propose three variations of transforming RDF into PG optionally in consideration of schemas: simple, generic, and complete. The authors formally show that two of the proposed mapping approaches (generic and complete) satisfy the property of information preservation, i.e., there exist inverse mappings that allow recovering the original dataset without information loss. 
The evaluation in this paper (Section~\ref{sec:experiments}) includes the schema-independent mapping referred to as the \emph{Generic Database Mapping} using the authors' implementation (RDF2PG\footnote{\scriptsize \label{fn:rdf2pg} \url{https://github.com/renzoar/rdf2pg}}). Although our focus is on RDF-star (instead of RDF), we include this approach since it provides the basic formalities and implementation that can be extended to support RDF-star.

In the opposite direction, Bruyat et al.~\cite{bruyat2021prec} propose PREC\footnote{\scriptsize \url{https://bruju.github.io/PREC/}}, a library that enables tranformation PGs into RDF graphs.
The authors built a uniform graph model to describe the structure of the property graph in RDF terms. PREC uses a \emph{context} in RDF-star format that describes the mappings between the terms used in the PG model and IRIs. The user can define a template representing the different properties and edges in the resulting RDF graph. Despite using RDF-star internally, the approach does not support mapping PGs to RDF-star graphs.

\paragraph{RDF-star and PG. }
Hartig et al.~\cite{RDFReconciliation} propose two approaches for transforming RDF-star to PG. 
The first approach maps ordinary RDF triples to edges in the PG. Metadata triples are then represented as edge properties. 
Our analysis (Section~\ref{sec:experiments}) includes this approach using the authors' implementation (RDF-star Tools\footnote{\scriptsize \label{fn:rdf-star-tools} \url{https://github.com/RDFstar/RDFstarTools}}). 
The second approach treats datatype and object property statements differently. The former are transformed into node properties and the latter into edges. This approach, however, is limited in mapping embedded triples, and it is not implemented in the RDF-star Tools library. 

Neosemantics is a well-known project to import RDF data into Neo4J, implemented as a Neo4j plug-in\footnote{\scriptsize \label{fn:newsemantics} Neosemantics: \url{https://neo4j.com/labs/neosemantics/}}. 
The implementation was only recently extended to include an RDF-star importing feature. 
As we will see in Section~\ref{sec:experiments} importing RDF-star into PG using this transformation is lossy and does not cover all cases. 

In the other direction, Khayatbashi et al.~\cite{khayatbashi2022converting} present an analysis evaluating three transformation approaches from PG to RDF, including an RDF-star approach. As a part of the study, the authors evaluated the performance of querying the generated RDF graphs in multiple triple stores. They found that there is no clear best mapping in terms of execution time; the performance of the queries over RDF and RDF-star graphs resulting from the mapping varies compared to their equivalent pure RDF representations. 

%% file: transformation-new.tex

Analyzing the approaches discussed in Section~\ref{sec:relatedWork}, we can extrapolate two principle approaches: 
\emph{RDF-topology Preserving Transformation (RPT)} and \emph{Property Graph Transformation (PGT)}. 
RPT tries to preserve the RDF-star graph structure by transforming each RDF statement into an edge in the PG. 
PGT, on the other hand, ensures that datatype property statements are mapped to node properties in the PG. 
In what follows, we first explain how these approaches transform RDF triples into PG and afterwards how the basic algorithms can be extended to support RDF-star. 


\begin{lstlisting}[
style={st:rdf},
caption={RDF triples},
label={lst:input}
]
@prefix ex: <http://example.org/> .
@prefix xsd: <http://www.w3.org/2001/XMLSchema#> .
ex:book  ex:publish_date "1963-03-22"^^xsd:date .
ex:book  ex:pages        "100"^^xsd:integer .
ex:book  ex:cover        20 .
ex:book  ex:index        "55" .
\end{lstlisting}

\begin{figure*}[htb]
%
\makebox[\linewidth][c]{
\begin{subfigure}[b]{.4\textwidth}
\centering
\includegraphics[width=.95\textwidth]{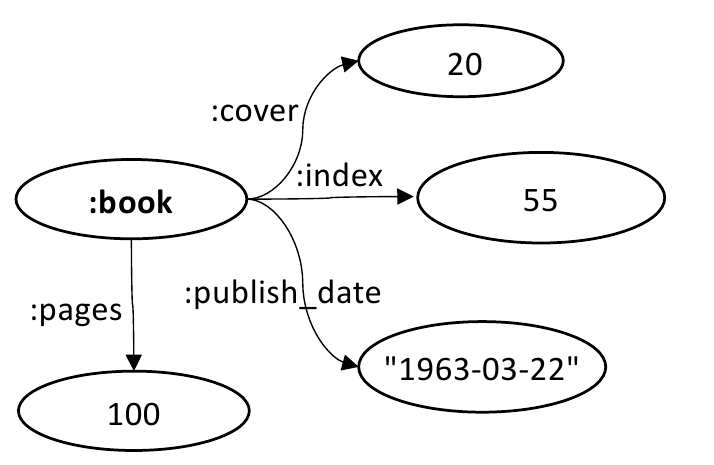}
\caption{\textbf{RTP}}
\end{subfigure}
\begin{subfigure}[b]{.4\textwidth}
\centering
\includegraphics[width=.95\textwidth]{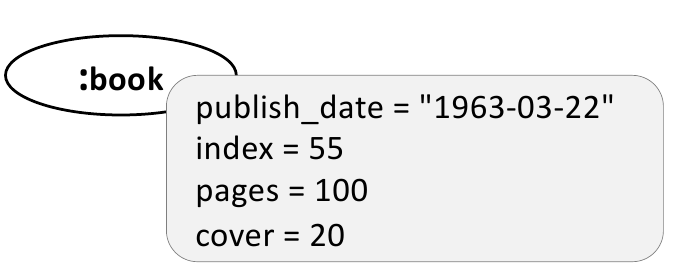}
\caption{\textbf{PGT}}
\end{subfigure}%
}
\caption{RTP and PGT transformations for the example in Listing~\ref{lst:input}}
\label{fig:MappingTypes}
\vspace*{-3ex}
\end{figure*}

Consider the example in Listing~\ref{lst:input} with multiple \emph{datatype property} statements describing the RDF resource \url{(ex:book)}. Figure~\ref{fig:MappingTypes} shows graphical visualizations of the property graphs generated by the two approaches: RPT (a) and PGT (b). 

RPT, for example, converts the triple \url{(ex:book, ex:index, 55)} into two nodes \url{(ex:book)} and \url{(55)}, connected by an edge \url{(ex:index)}. All other triples involving RDF resources, blank nodes, or literal values can be transformed in a similar way so that we obtain the PG in Figure~\ref{fig:MappingTypes}(a). 
Algorithm~\ref{alg:rdfTrdf} formalizes the RPT approach; for each triple it always creates a node for the subject (line 3) and the object (line 5) with an edge connecting them (line 12) -- of course avoiding duplicate nodes for the same IRIs.


For the same example (Listing~\ref{lst:input}), PGT creates the PG in Figure~\ref{fig:MappingTypes}(b) consisting of a single node respresenting the RDF resource \url{(ex:book)} with multiple properties representing property-object pairs from the RDF statements, such as \url{(ex:index, 55)}. 
Distinguishing between datatype and object property statements, this approach transforms object property statements to edges and datatype property statements to properties of the node representing the subject. 
Unlike RPT, the resulting PG nodes represent only RDF resources or blank nodes while literal objects will become properties. 
PGT is more formally sketched in Algorithm~\ref{alg:pgTrdf}, which first checks the type of the statement's object (line 5) and based on that decides to either create a node (if it does not yet exist, line 6) or a property (line 13). 

\savebox{\algleft}{%
\begin{minipage}{.5\textwidth}
\vspace*{-2.5ex}
\begin{algorithm}[H]
\caption{RPT}
\label{alg:rdfTrdf}
\begin{scriptsize}
\begin{algorithmic}[1]
\REQUIRE A set of RDF Triples $T$ 
\ENSURE A Property Graph $Pg =  (N, E, edge, lbl, P, \sigma)$
\STATE $Pg \leftarrow \emptyset $
\FOR{$t \in T$, such that $t = <s,p,o>$}
\STATE {$N= N \cup$ \{$s$\} }
 \STATE {$lbl(s)=$ \{"RDF resource"\}}
     \STATE {$N= N$ $\cup$ \{$o$\}}
 \IF{$o$ is an RDF resource}
         \STATE {$lbl(o)=$ \{"RDF resource"\}}
        \ELSE
         \STATE {$lbl(o)=$ \{"Literal"\}}
        \ENDIF
        \STATE {$E= E$ $\cup$ \{$e$\}}
        \STATE { $edge(e)= (s,o) $}
        \STATE {$lbl(e)= p $}
    \ENDFOR
    \RETURN $Pg$
\end{algorithmic}
\end{scriptsize}
\end{algorithm}
\vspace*{-2.3ex}
\end{minipage}}

\savebox{\algright}{%
\begin{minipage}{.5\textwidth}
\vspace*{-2.6ex}
 \begin{algorithm}[H]
 \caption{PGT}
\label{alg:pgTrdf}
\begin{scriptsize}
\begin{algorithmic}[1]
\REQUIRE A set of RDF Triples $T$
\ENSURE A Property Graph $Pg =  (N, E, edge, lbl, P, \sigma)$
\STATE $Pg \leftarrow \emptyset $
\FOR{$t \in T$, such that $t = <s,p,o>$}
\STATE {$N= N$ $\cup$ \{$s$\} }
 \STATE {$lbl(s)=$ \{"RDF resource"\}}
 \IF{$o$ is an RDF resource}
     \STATE {$N= N$ $\cup$ \{$o$\}}
      \STATE {$lbl(o)=$ \{"RDF resource"\}}
       \STATE {$E= E$ $\cup$ \{$e$\}}
        \STATE { $edge(e)= (s,o) $}
        \STATE { $lbl(e)= p $}
        \ELSE
       \STATE {$P$= $P$ $\cup$ \{$pr$\}}
        \STATE  {$pr = $ \{(p, o)\} }
        \STATE {$\sigma(s)=$ pr}
 \ENDIF
    \ENDFOR
     \RETURN $Pg$
\end{algorithmic}
\end{scriptsize}
\end{algorithm}
\vspace*{-1.7ex}
\end{minipage}}

\noindent\usebox{\algleft}\hfill\usebox{\algright}%
\medskip 

\begin{figure*}[htb]
\vspace*{-3ex}
\makebox[\linewidth][c]{
\begin{subfigure}[b]{.5\textwidth}
\centering
\includegraphics[width=.95\textwidth]{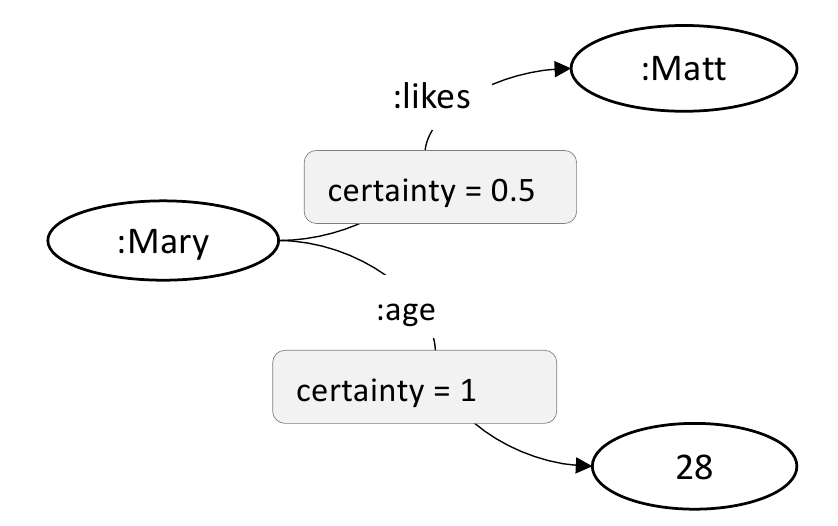}
\caption{\textbf{RTP}}
\end{subfigure}
\begin{subfigure}[b]{.5\textwidth}
\centering
\includegraphics[width=.95\textwidth]{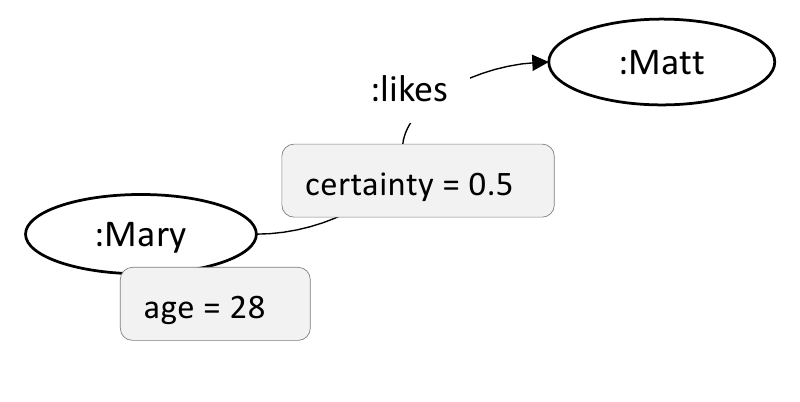}
\caption{\textbf{PGT}}
\end{subfigure}%
}
\caption{RPT and PGT transformations for the example in Listing~\ref{lst:inputstar}}
\label{fig:MappingTypesstarfail}
\vspace*{-3ex}
\end{figure*}

Let us now consider the RDF-star example in Listing~\ref{lst:inputstar}, which contains an asserted triple for an embedded data property statement -- the PGs obtained by applying RTP and PGT are shown in Figure~\ref{fig:MappingTypesstarfail}. Algorithms~\ref{alg:RPTstar} and~\ref{alg:PGTstar} illustrate the main principle of mapping the embedded and asserted triples. RPT conversion for RDF-star is identical to RDF triples, then converting the asserted triple into an edge property (Algorithm~\ref{alg:RPTstar} lines 5-8). PGT transforms the embedded triple depending on its object; if it is an RDF resource, PGT converts it to an edge. Otherwise, it converts the embedded triple into a node with a property (Algorithm~\ref{alg:PGTstar} lines 6-11) and fails to transform the asserted triple.   
\vspace*{-0.3ex}

In summary, the transformation of the triples from Listing~\ref{lst:inputstar} using PGT results in a PG with a single node that makes it impossible to represent the asserted triple since PGs do not support properties over other properties. In contrast, RPT transforms the embedded triple into an edge in the PG and can express the asserted triple as the edge's property. 
The Abstracting away from a few details (see also Section~\ref{sec:experiments}), the Neosemantics approach\footref{fn:newsemantics} basically follows PGT while RDF-star Tools\footref{fn:rdf-star-tools} and RDF2PG\footref{fn:rdf2pg} follow RPT.

\savebox{\algleft}{%
\begin{minipage}{.49\textwidth}
\vspace*{-2.5ex}
\begin{algorithm}[H]
\caption{RPT-star}
\label{alg:RPTstar}
\begin{scriptsize}
\begin{algorithmic}[1]
\REQUIRE  A set of RDF-star Triples $T$ 
\ENSURE A Property Graph $Pg = (N, E, edge, lbl, P, \sigma)$
\STATE $Pg \leftarrow \emptyset $
\FOR{$t \in T$, such that $t = <s,p,o>$}
 \IF{$\bar{t}$ is an embedded triple, such that $t = <\bar{t},p,o>$ and $\bar{t}= <\bar{s},\bar{p},\bar{o}>$ } 
   \STATE {$PgOut =$ RPT($\bar{t}$)} 
     \STATE  {$pr =$ \{(p, o)\} }
    \STATE {$P$ = $P$ $\cup$ \{$pr$\}}
    \STATE {$\sigma(e)= pr $}
     \ENDIF
         \ENDFOR
       \RETURN $Pg$ $\cup$ $PgOut$
\end{algorithmic}
\end{scriptsize}
\end{algorithm}
\vspace*{-3ex}
\end{minipage}}

\savebox{\algright}{%
\begin{minipage}{.49\textwidth}
 \begin{algorithm}[H]
 \caption{PGT-star}
\label{alg:PGTstar}
\begin{scriptsize}
\begin{algorithmic}[1]
\REQUIRE A set of RDF-star Triples $T$
\ENSURE A Property Graph $Pg =  (N, E, edge, lbl, P, \sigma)$
\STATE $Pg \leftarrow \emptyset $
\FOR{$t \in T$, such that $t = <s,p,o>$}
 \IF{$\bar{t}$ is an embedded triple, such that $t = <\bar{t},p,o>$ and $\bar{t}= <\bar{s},\bar{p},\bar{o}>$ } 
 
\IF{$\bar{o}$ is an RDF resource}
 \STATE {$PgOut$ =  PGT($\bar{t}$)} 
        \ELSE
        \STATE {$PgOut$ =  PGT($\bar{t}$)} 
        \STATE {$pr = $ \{($\bar{p}$,$\bar{o}$)\}}
        \STATE {$P = P$ $\cup$ \{$pr$\}}
        \STATE {$\sigma(\bar{s})= pr$}
          
        \ENDIF
 \ENDIF
    \ENDFOR
     \RETURN $Pg$ $\cup$ $PgOut$
\end{algorithmic}
\end{scriptsize}
\end{algorithm}
\vspace*{-3ex}
\end{minipage}}

\noindent\usebox{\algleft}\hfill\usebox{\algright}%
\begin{lstlisting}[
style={st:rdf},
caption={RDF-star triples},
label={lst:inputstar}
]
@prefix ex: <http://example.org/> .
<<ex:Mark ex:age 28>> ex:certainty 1 .
\end{lstlisting}

%% file: casesws.tex

In this section, we present a systematic list of test cases that transformation approaches need to fulfill. We distinguish between basic cases that conform to small RDF graphs as well as a range of RDF-star specific test cases that challenge existing approaches -- as we will see in our evaluation in Section~\ref{sec:experiments}. 
The complete list of test cases with their short titles is shown in Table~\ref{tab:cases} -- subcases represent variations and bold font indicates cases discussed in more detail in this paper, more details for all cases are available on our website\footnote{\scriptsize \label{fn:website} \url{https://relweb.cs.aau.dk/rdfstar}} and Appendix~\ref{App:cases}. 


\begin{table*}[htb]
  \caption{Test cases for evaluating RDFstar-to-PG transformation approaches}
  \label{tab:cases}
  \centering
\scalebox{0.95}{
  \begin{tabular}{ll}
\toprule
\multicolumn{2}{c}{\textit{Standard RDF}} \\
 \midrule
      Case & Description \\
 \textbf{1}		& \textbf{Standard RDF statement} \\  
 \textbf{2}	& \textbf{The predicate of an RDF statement is subject in another statement}\\ 
 2.1	& Predicate as subject and literal as object \\
 2.2	& Predicate as subject and RDF resource as object\\ 
 2.3	& Predicate as subject and RDF property as object - \emph{rdfs:subPropertyOf}\\
 2.4	& Predicate as subject and RDF class as object - \emph{rdf:type} \\ 
 \textbf{3} & \textbf{Data types and language tags}\\
 3.1 & Datatype property statements with different data types of the literal objects \\
 3.2 & Datatype property statements with different language tags of the literal objects \\
 {4}	& {RDF list}\\
 {5}	& {Blank nodes}\\
 {6}	& {Named graphs}\\
 {7}	& {Multiple types for resources - \emph{rdf:type}}\\
 \midrule 
\multicolumn{2}{c}{\textit{RDF-Star}} \\
\midrule
 \textbf{8}	& \textbf{Embedded \emph{object property} statement in subject position} \\ 
 \textbf{9}	& \textbf{Embedded \emph{datatype property} statement in subject position} \\
 \textbf{10} & \textbf{Embedded \emph{object property}statement in object position} \\
{11}	& {Embedded \emph{object property} statement in subject position and non-literal object} \\ 
 11.1 & {Asserted statement with non-literal object} \\
 11.2	& {Asserted statement with non-literal object that appears in another asserted statement} \\
 {12} & {Embedded statement in subject position - \emph{object property} with \emph{rdf:type} predicate} \\ 
 12.1 & {Asserted statement with \emph{rdf:type} as predicate} \\
 12.2	& {Embedded statement with \emph{rdf:type} as predicate} \\
 {13} & {Double nested RDF-star statement in subject position} \\ 
 {14} & {Multi-valued properties}\\ 
 14.1	& {RDF statements with same subject and predicate and different objects} \\
 14.2	& {RDF-star statements with the same subject and predicate and different objects} \\
 {15} & {Multiple instances of embedded statements in a single RDF-star graph} \\ 
 15.1 & {Identical embedded RDF-star statements with different asserted statements } \\
 15.2	& {RDF statements as embedded and asserted statements in the same graph} \\
\bottomrule
\end{tabular}
}
\end{table*}

\subsection{Standard RDF}
This list of test cases (1-7) is related to the basic features of the RDF model, such as object and data type property RDF statements, data types, blank nodes, named graphs, and others. This section presents some examples of cases (1-3) and their variations. The complete list of cases can be found in Section~\ref{App:rdf} of Appendix~\ref{App:cases}. 

\stitle{Case 1: Standard RDF statement}
This case represents an \emph{object property} statement. Both, subject and object are RDF resources. Most transformation approaches map this case to two nodes (subject and object) with an edge (the predicate) connecting them. 
\begin{lstlisting}[
style={st:rdf},
label={lst:case1}
]
@prefix ex: <http://example.org/> .
ex:alice ex:meets ex:bob .
\end{lstlisting}

\stitle{Case 2: The predicate of an RDF statement is subject in another statement}
Mapping an RDF statement to two nodes with the predicate as label of the edge between them leads to problems when the predicate itself is also used as a subject in another RDF statement -- Case 2.1 therefore consists of the following statements: 
\begin{lstlisting}[
style={st:rdf},
label={lst:Edge2Edge}
]
@prefix rdfs: <http://www.w3.org/2000/01/rdf-schema#> .
@prefix ex: <http://example.org/> .
ex:Sam ex:mentor ex:Lee .
ex:mentor rdfs:label "project supervisor" .
ex:mentor ex:name "mentor's name" .
\end{lstlisting}
Other variants of Case 2 include a predicate os object as well as \texttt{rdf:type} and \texttt{rdfs:subPropertyOf}. 

\stitle{Case 3: Data types and language tags} 
It is also important to test the support of different data types and language tags. 
Hence, Case 3.1, for instance,  contains several \emph{datatype property} statements involving different data types and formats for the literal objects:
\begin{lstlisting}[
style={st:rdf},
label={lst:Case3.1}
]
@prefix ex: <http://example.org/> .
@prefix xsd: <http://www.w3.org/2001/XMLSchema#> .
ex:book  ex:publish_date "1963-03-22"^^xsd:date .
ex:book  ex:pages        "100"^^xsd:integer .
ex:book  ex:cover        20 .
ex:book  ex:index        "55" .
\end{lstlisting}
\subsection{RDF-star}

In what follows, we present test cases that exemplify different RDF-star features. The test cases (8-15) and their variations are 12 cases in total (see Table~\ref{tab:cases}), and we present some examples from Cases 8, 9, and 10. The complete list of cases can be found in Section~\ref{App:rdfs} of Appendix~\ref{App:cases}. 

\stitle{Case 8: Embedded object property statement in subject position} 
As the name indicates and the following listing shows, this test case features an RDF-star statement where the subject corresponds to an embedded object property statement and the object is a literal: 
\begin{lstlisting}[
style={st:rdf},
label={lst:rdfstarexample}
]
@prefix ex: <http://example.org/> .
<<ex:alice ex:likes ex:bob>> ex:certainty 0.5 .
\end{lstlisting}

\stitle{Case 9: Embedded datatype property statement in subject position} 
Similar to the previous case we again have an RDF-star statement where the subject corresponds to an embedded statement. In contrast to the Case 8, the embedded statement in Case 9 is a datatype property statement:
\begin{lstlisting}[
style={st:rdf},
label={lst:case4}
]
@prefix ex: <http://example.org/> .
<<ex:Mark ex:age 28>> ex:certainty 1 .
\end{lstlisting}

\stitle{Case 10: Embedded object property statement in object position} 
Of course, RDF-star statements can also have embedded statements on object position, which is covered in this case. Similar to Case 8, the embedded statement is an \emph{object property} statement. 
\begin{lstlisting}[
style={st:rdf},
label={lst:case5}
]
@prefix ex: <http://example.org/> .
ex:bobhomepage ex:source <<ex:mainPage ex:writer ex:alice>> .
\end{lstlisting}

\noindent
Other test cases cover other variations of asserted statements (Case 11), the usage of \texttt{rdf:type} in the embedded and asserted statements (Case 12), the double nesting of RDF-star statements (Case 13), the same RDF-star statement with different asserted statements (Case 14), and multiple occurrences of an RDF-star statement within the same graph (Case 15). As mentioned above, details can be found on our project website\footref{fn:website}. 

%% file: discussionws.tex

In this section, we use the test cases identified in Section~\ref{sec:cases} to evaluate a number of transformation approaches that we have identified in Section~\ref{sec:relatedWork}: RDF2PG\footref{fn:rdf2pg}, RDF-Star Tools\footref{fn:rdf-star-tools}, and Neosemantics~\footref{fn:newsemantics}. We use the property graph model introduced in Section~\ref{sec:preliminaries} to represent the property graph generated by each approach. The complete results can be found in Table~\ref{tab:mapping} in Appendix~\ref{App:results}. As we will see and as already mentioned in Section~\ref{sec:transformation}, RDF-star Tools and RDF2PG follow the RDF-topology Preservation Transformation (RPT) whereas Neosemantics adopts the Property Graph Transformation (PGT). 



\begin{figure*}[thb]
\makebox[\linewidth][c]{%
\begin{subfigure}[b]{.5\textwidth}
\centering
\includegraphics[width=.95\textwidth]{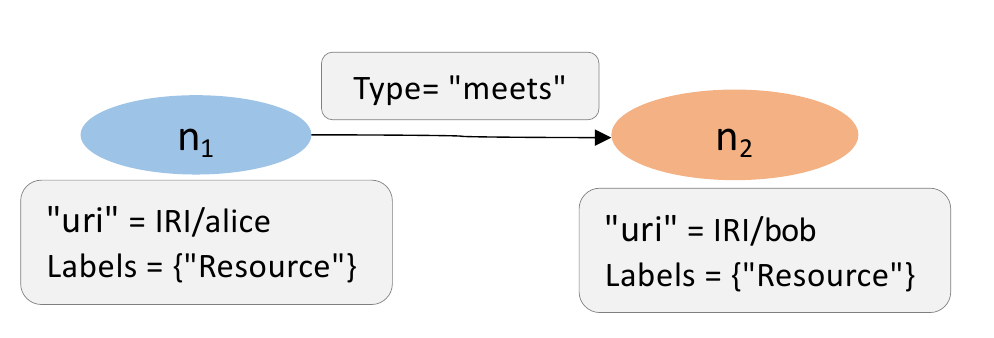}
\caption{\textbf{Neosemantics}}
\end{subfigure}%
\begin{subfigure}[b]{.5\textwidth}
\centering
\includegraphics[width=.95\textwidth]{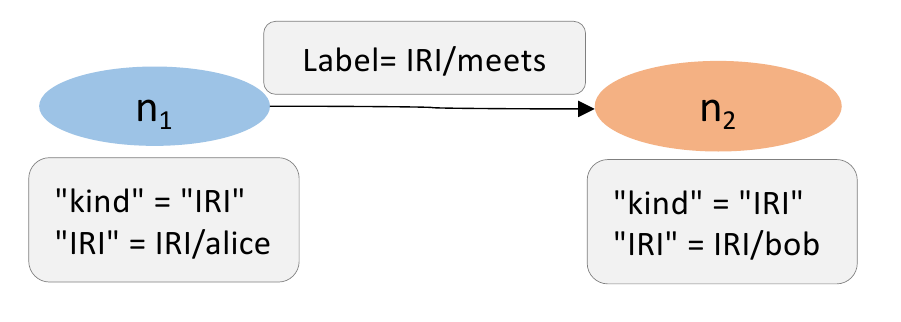}
\caption{\textbf{RDF-star Tools}}
\end{subfigure}
\begin{subfigure}[b]{.5\textwidth}
\centering
\includegraphics[width=.95\textwidth]{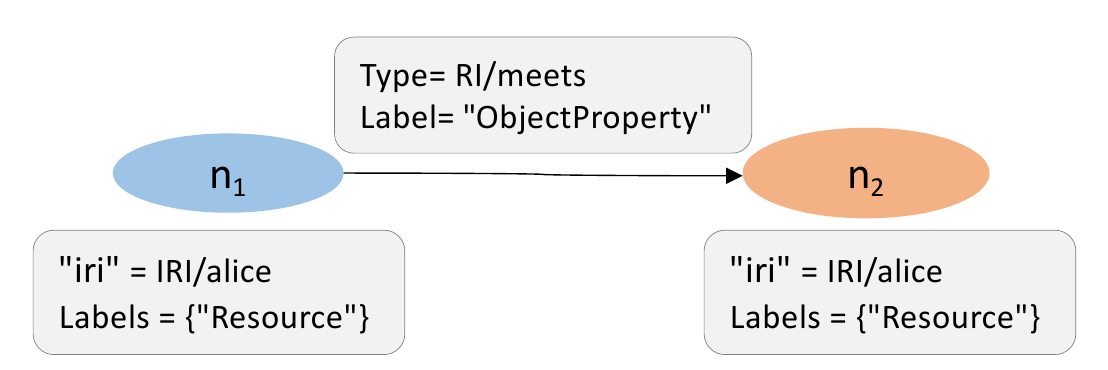}
\caption{\textbf{RDF2PG}}
\end{subfigure}
}
\makebox[\linewidth][c]{%
\begin{subfigure}[b]{.5\textwidth}
\includegraphics[width=.95\textwidth]{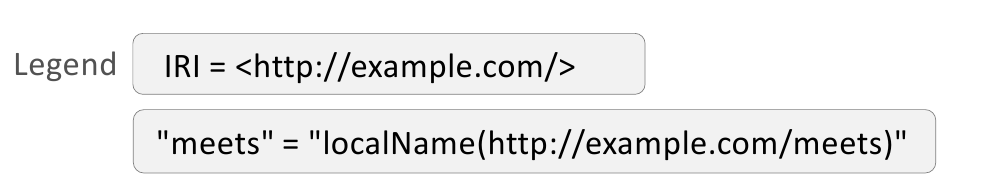}
\end{subfigure}

}
\caption{PGs obtained for Case 1}
\label{fig:objectStatementResults}
\end{figure*}

\subsection{Standard RDF}

At first, let us discuss our findings for Cases 1 through 7 from Table~\ref{tab:cases} targeting standard RDF statements.
Case 1 corresponds to a simple RDF statement with IRIs as subject and object. Non-surprisingly, all three libraries create a PG with two nodes and one edge (see  Figure~\ref{fig:objectStatementResults}). 
The main differences are in the way types, labels, and properties are handled.
Both Neosemantics and RDF2PG use ``Resource'' as the label of the two nodes and a key value pair (key=IRI, value=IRI of the subject/object). Additionally, RDF-star Tools uses two additional properties for the nodes: (key=``kind'', value=``IRI'') and (key=``IRI'', value=subject/object IRI). 
Whereas Neosemantics and RDF2PG use the predicate from the RDF statement as the edge's type, RDF-star Tools uses the predicate as an edge label. RDF2PG additionally uses ``ObjectProperty'' as an additional edge label. 



\begin{figure*}[htb]
\makebox[\linewidth][c]{%
\begin{subfigure}[b]{.5\textwidth}
\centering
\includegraphics[width=.92\textwidth]{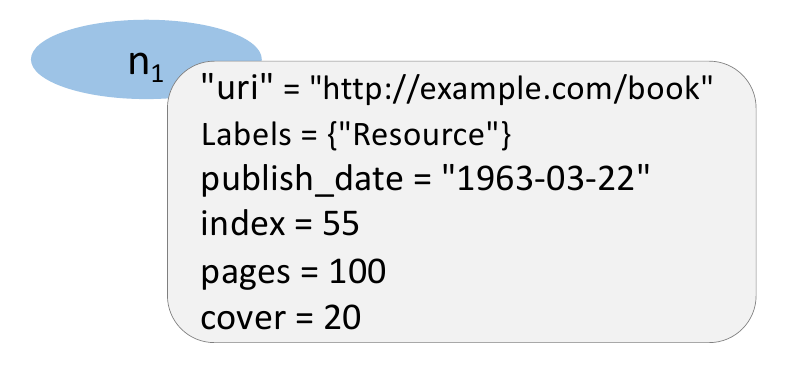}
\caption{\textbf{Neosemantics}}
\end{subfigure}
\begin{subfigure}[b]{.7\textwidth}
\centering
\includegraphics[width=.99\textwidth]{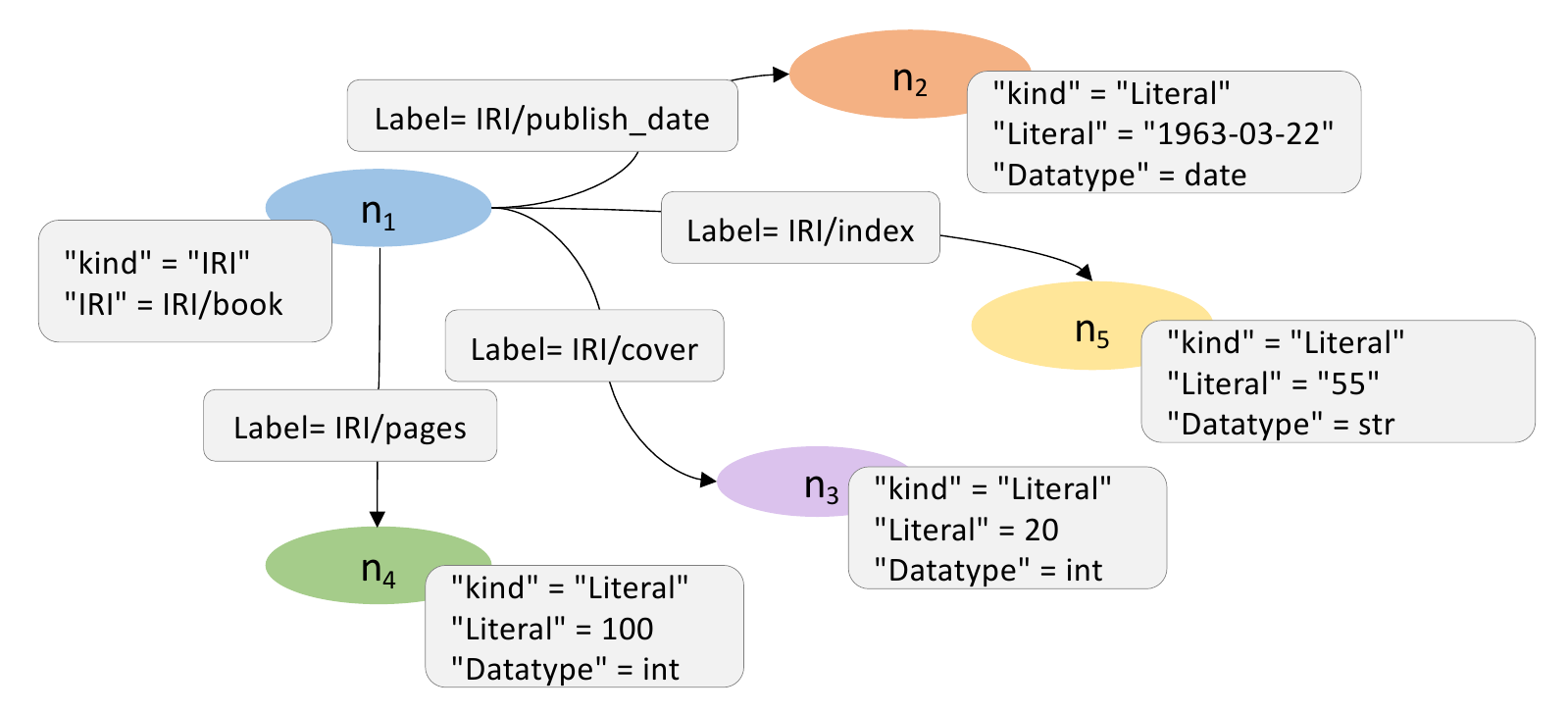}
\caption{\textbf{RDF-star Tools}}
\end{subfigure}%
}\\
\makebox[\linewidth][c]{%
\begin{subfigure}[b]{.7\textwidth}
\centering
\includegraphics[width=.99\textwidth]{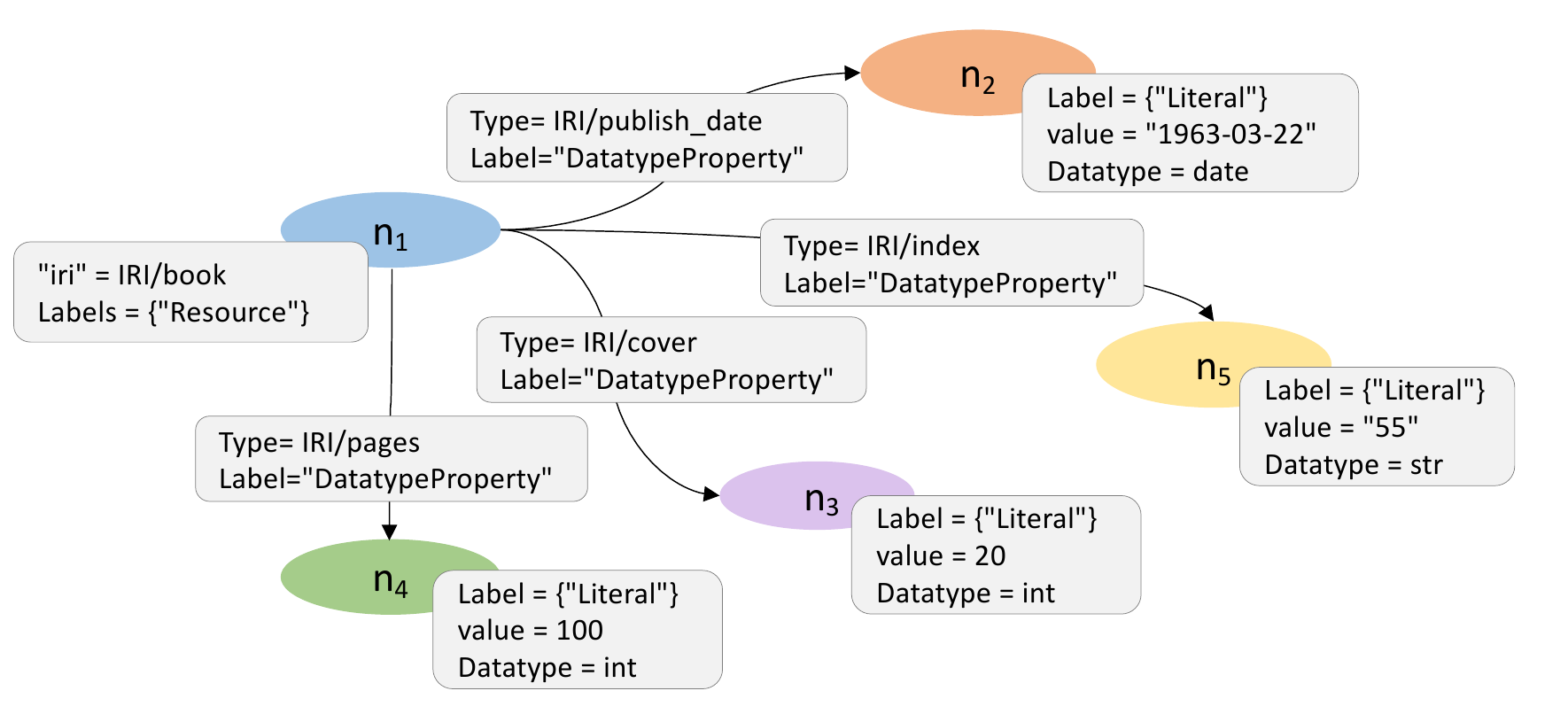}
\caption{\textbf{RDF2PG}}
\end{subfigure}
}
\caption{PGs obtained for Case 3}
\label{fig:edgesvsproperty}
\end{figure*}

Case 3.1 tests the transformation of \emph{datatype property} statements.
In this case, the differences between the libraries are more evident, as depicted in Figure \ref{fig:edgesvsproperty}. 
RDF-star Tools and RDF2PG represent each literal using a separate node- 
The properties of the edge are similar to the ones described for Case 1 with RDF2PG using ``DatatypeProperty'' as edge label.
Neosemantics instead, only creates a single node representing the complete set of input RDF statements; the nodes has one property for each datatype property statement in the input.
%
Additionally, Cases 3.1 and 3.2 also test how transformation approaches support data types and language tags in RDF statements. While Neosemantics ignores them, both RDF-star Tools and RDF2PG define the nodes representing the literal objects with type literal and use the XSD schema data type as annotation.


Cases 4--6 test different RDF features, such as RDF lists (case 4), blank nodes (case 5), and named graphs (case 6). All three projects support RDF lists and blank nodes but only Neosemantics also supports namged graphs. 

\begin{figure*}[htb]
\makebox[\linewidth][c]{%
\begin{subfigure}[b]{.5\textwidth}
\centering
\includegraphics[width=.75\textwidth]{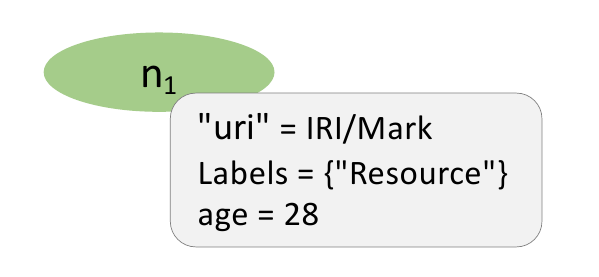}
\caption{\textbf{Neosemantics}}
\end{subfigure}
\begin{subfigure}[b]{.7\textwidth}
\centering
\includegraphics[width=.92\textwidth]{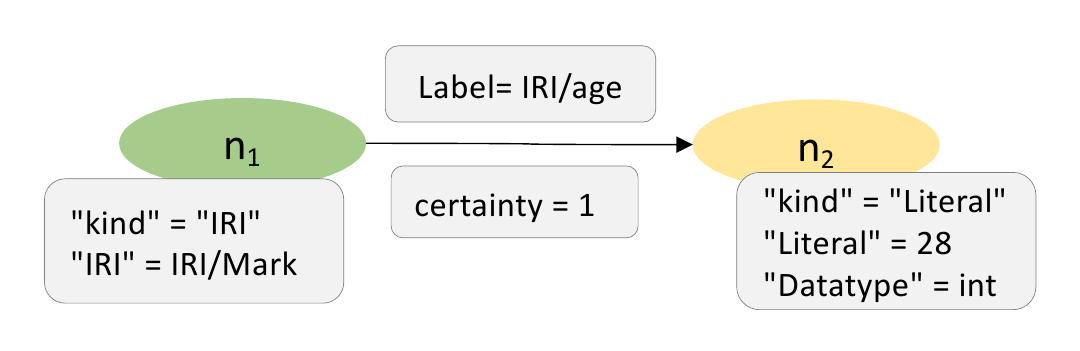}
\caption{\textbf{RDF-star Tools}}
\end{subfigure}%
}
\caption{PGs obtained for Case 9}
\label{fig:rdfconverttpes}
\end{figure*}

\subsection{RDF-star}
Let us now discuss over findings for evaluating Cases 8 through 13 from Table~\ref{tab:cases} targeting diverse RDF-star constructs. We have observed that many of them are not supported by existing libraries. 
RDF2PG does not support embedded RDF-star statements, so none of these cases could be converted to a PG. 


Neosemantics ignores RDF-star statements where:
\begin{inparaenum}[(i)]
    \item the object of the asserted statement is a literal value (Case 9) or an embedded statement (Case 10), 
    \item all elements of an RDF-star statement are RDF resources (Cases 11.1 and 11.2), 
    \item the predicate of the embedded or asserted statement is "rdf:type" (Cases 12.1 and 12.2), or 
    \item the RDF-star statement is a double embedded statement (Case 13). 
\end{inparaenum}

RDF-star Tools supports most cases by creating nodes in the PG for literal objects. 
However, Cases 10 and 13 cause an error in the reading phase of the conversion: we believe that this is an implementation error, and both cases could be supported in the same manner as the others. 

Among all cases, Case 9 is particularly interesting and represents a natural extension of Case 3.1.
If the embedded statement \url{(mark, age, 25)} is translated in PGT-style, e.g., by Neosemantics, \url{mark} is represented as a node with a property having \url{age} as key and \url{25} as value. It is then not straightforward how to convert the asserted statement (stating that the certainty of \url{(mark, age, 25)} is \url{1}). Hence, when running Neosemantics, we noticed that it does not transform the asserted statement completely. 
RDF-star Tools (RPT-style transformation), on the other hand, converts the statement in Case 9 by transforming the embedded statement to an edge with type \url{age} between two nodes representing \url{mark} and \url{25}; the part of the asserted statement is represented as a key-value property associated to the \url{age} edge. 

\subsection{Edge cases}
Case 2 (resource both as a predicate and a subject/object in another statement) is not supported correctly by existing transformation approaches. The result PGs include two separate and independent elements for such resources: one node and one edge property. This is different from RDF, where an IRI identifies the same resource independently from its position in an RDF statement.

Annotating RDF predicates (both datatype and object properties) is a common way to model schema information in RDF.
Our finding suggests that there may be more conversion problems when looking at RDF graphs that include RDF Schema and OWL axioms.
This investigation is beyond the scope of this paper but we plan to investigate it in our future research.

Other interesting cases are those where the RDF input contains multiple statements with the same subject and predicate but different objects. 
In Case 14.1, for example, contains two datatype property statements with the same subjects and predicates. 
Neosemantics converts the literals \url{("Info\_page", "aau\_page")} into a list of strings\footnote{\scriptsize \url{https://neo4j.com/docs/cypher-manual/current/syntax/values/\#composite-types}} used as a value in a property of the node representing the subject of the RDF statement. 

\subsection{RDF-topology Preservation Transformation (RPT) and Property Graph Transformation (PGT)}
Our analysis confirmed our observation that RDF-star Tools follows the RPT approach; the approach creates a node for each distinct subject and object of every RDF statement. Predicates and objects of RDF-star statements are transformed into properties with the key corresponding to the predicate and the value to the object. 
Neosemantics, in contrast, follows PGT approach for most of the datatype property statements, e.g., Cases 2.1, 3.1, 3.2, 11.2, and 14.1, and supports types and labels (Case 2.4, 5, 6, and 7) slightly differently. 
Finally, the RDF2PG project adopts the RPT approach in all cases handled.

In summary, the two lines of transformation approaches, RPT and PGT, form the core of start-of-the-art libraries. However, none of the tested libraries supports all test cases.
In the end, the particular requirements for converting RDF-star to PGs vary based on the particular use cases and application-specific needs. 
In one use case, a user might favor one transformation approach over the other based on ontology availability, the application domain, performance, or additional application-specific needs. 
And driven by real-world applications, users might want to adopt a combination of the basic transformation approaches to best fit their needs.

%% file: approachws.tex


\stitle{Ontology and schema availability} 
Aapplication domains, such as financial services, life science, and military, have well-established domain ontologies that are native to the RDF/RDF-star model.
Such use cases often also exhibit computationally expensive queries and algorithms that are better for PGs, 
Hence, in such cases, RPT is preferable to convert the ontology and PGT to convert instance data to reduce the number of nodes and therefore improve runtime. 

Another use case involves harvesting Linked Open Data~\cite{linked2014,HartigHS19} where the PGT approach is usually preferable. Likewise, PGT is easier to use in combination with tabular data. 
However, in situations, where the ontology is essential and the RDF data is highly complex and heterogeneous, users may prefer RPT. In the end, some graph-based machine learning models also require graphs that model literal property values as nodes.

\stitle{Performance and query complexity} 
Query performance over graphs generally depends on the number of edges and nodes~\cite{das2014tale}. Therefore, the user may choose to convert an RDF/RDF-star graph into a PG using a specific transformation method to comply with performance requirements. For example, in an application for a smart home, the authors transformed the RDF graph into a PG based on a custom transformation~\cite{baken2020linked}. 
The authors evaluated the usage of a PG generated by the custom transformation versus the PG generated by NSMTX\footnote{\scriptsize \url{https://neo4j.com/nsmtx-rdf/}}. The PG of the custom transformation had two nodes and one edge compared to 16 nodes and 15 edges for the PG generated by the NSMTX plugin. 
As a result, executing queries over the former graph is more efficient than over the NSMTX graph. 
The custom transformation eliminated many edges (i.e., relations) and converted them into properties. 
Since RPT tends to convert each triple into an edge, the transformation generates many nodes compared to PGT that wraps some triples as properties for nodes. 

\stitle{Data sharing} 
RPT allows RDF-star asserted triples (i.e., edge attributes) to be represented as nodes in the PG, treating them as individual entities. This approach can make the graph representation more expressive for other users to understand and re-use than PGT. 
On the other hand, the PGT can result in many nodes with properties as literal key-value pairs, not explicit edges. This representation can be a natural choice to represent descriptions of nodes.

%% file: conclusion.tex

In this paper, we have evaluated and discussed how to transform RDF-star graphs into property graphs. To evaluate existing approaches (Neosemantics, RDF-star Tools, and RDF2PG), we have identified a number of test cases. 
Our analysis has shown that none of these three approaches supports all test cases and that none of them is the best for all applications.
None of them considers user requirements for the transformation process. 
Nevertheless, existing approaches can roughly be categorized into two lines of transformation approaches: RDF-topology preserving transformation (RPT) and PG transformation (PGT).
In our future work, we plan to expand our experiments to comprehensive datasets and combining RPT and PGT into a single hybrid transformation approach. 
Furthermore, based on our previous work, we plan to investigate how processing advanced queries over provenance-enhanced datasets can be improved~\cite{GalarragaJHP18,GalarragaMH17,HansenLGLTH20,LissandriniMHP22}, how querying heterogeneous provenance-enhanced graph data can be embedded in scalable ecosystems~\cite{SakrBVIAAAABBDV21}, how evolving knowledge graphs can be supported~\cite{Hose21}, and how interoperability between RDF-star and property graphs can be achieved~\cite{SagiLPH22}. 

%% file: cases.tex
This section contains the complete list of test cases, including the cases presented in Section~\ref{sec:cases}. The cases are used to evaluate the transformation approaches from RDF and RDF-star to property graphs. 
\subsection{Standard RDF}
\label{App:rdf}

\stitle{Case 1: Standard RDF statement}
This case represents an \emph{object property} statement. Both, subject and object are RDF resources. Most transformation approaches map this case to two nodes (subject and object) with an edge (the predicate) connecting them. 
\begin{lstlisting}[
style={st:rdf},
label={lst:case1}
]
#Case 1
@prefix ex: <http://example.org/> .
ex:alice ex:meets ex:bob .
\end{lstlisting}

\stitle{Case 2: The predicate of an RDF statement is subject in another statement}
Mapping an RDF statement to two nodes with the predicate as label of the edge between them leads to problems when the predicate itself is also used as a subject in another RDF statement as in Case 2.1 example. 
Particularly, in the RDF model, an IRI resource may become a subject in one triple and a predicate or an RDF property~\footnote{The concept of an RDF property defines as the relationship between subject and object resources. \emph{rdf:Property} is the class of RDF properties in the RDF schema~\cite{RDFS}} in another. Such a scenario is common as RDF syntax allows us to
make assertions about any resources, including predicates. For example, consider the triple (ex:mentor, rdfs:label, "project supervisor") in the RDF graph in Case 2.1. It includes metadata about the predicate, ex:mentor. More specifically, the predicate (ex:mentor) of the triple (ex:Sam, ex:mentor, ex:Lee) becomes a subject in the second triple (ex:mentor, rdfs:label, "project supervisor").
\begin{lstlisting}[
style={st:rdf},
label={lst:Edge2Edge}
]
#Case 2.1
@prefix rdfs: <http://www.w3.org/2000/01/rdf-schema#> .
@prefix ex: <http://example.org/> .
ex:Sam ex:mentor ex:Lee .
ex:mentor rdfs:label "project supervisor" .
ex:mentor ex:name "mentor's name" .
\end{lstlisting}
Other variations of this case can include the \emph{rdfs:subPropertyOf} predicate. 
The triple (ex:supervise, rdfs:subPropertyOf, ex:administer) is an example of this case (See Case 2.2). The RDF model allows metadata about properties, i.e., an RDF property is a \emph{subPropertyOf} of another RDF property (See Case 2.3). A different variation would be a triple where the RDF predicate becomes a subject with \emph{rdf:type} predicate such as (ex:friend, rdf:type, ex:relation) (See Case 2.4). That is common because the RDF graph can contain both data and schema (definition of RDF properties or RDF types). In such a case, it is necessary for the transformation method to decide whether to extract the schema (and transform it independently) or process the schema as part of the data.

\begin{lstlisting}[
style={st:rdf},
label={lst:subproperty}
]
#Case 2.2:
@prefix rdfs: <http://www.w3.org/2000/01/rdf-schema#> .
@prefix ex: <http://example.org/> .
ex:Martin ex:mentorJoe ex:Joe .
ex:mentorJoe ex:alias ex:teacher . 
\end{lstlisting}

\begin{lstlisting}[
style={st:rdf},
label={lst:subproperty}
]
#Case 2.3:
@prefix rdfs: <http://www.w3.org/2000/01/rdf-schema#> .
@prefix ex: <http://example.org/> .
ex:Jan ex:supervise ex:Leo .
ex:supervise rdfs:subPropertyOf ex:administer .
\end{lstlisting}

\begin{lstlisting}[
style={st:rdf},
label={lst:subproperty}
]
#Case 2.4:
@prefix rdfs: <http://www.w3.org/2000/01/rdf-schema#> .
@prefix ex: <http://example.org/> .
ex:Tom ex:friend ex:Chris .
ex:friend rdf:type ex:relation .
\end{lstlisting}
\stitle{Case 3: Data types and language tags} 
It is also important to test the support of different data types and language tags. Hence, Case 3.1, for instance,  contains several \emph{datatype property} statements involving different data types and formats for the literal objects: 
\begin{lstlisting}[
style={st:rdf},
label={lst:Case3.1}
]
#Case 3.1:
@prefix ex: <http://example.org/> .
@prefix xsd: <http://www.w3.org/2001/XMLSchema#> .
ex:book  ex:publish_date "1963-03-22"^^xsd:date .
ex:book  ex:pages        "100"^^xsd:integer .
ex:book  ex:cover        20 .
ex:book  ex:index        "55" .
\end{lstlisting}
In Case 3.2 example, a list of RDF triples that have literal values as objects. The literal objects contain language tag. 
\begin{lstlisting}[
style={st:rdf},
label={lst:Case3.2}
]
#Case 3.2:
@prefix ex: <http://example.org/> .
ex:book  ex:Englishtitle "Book"@en .
ex:book  ex:title "Bog"@da .
\end{lstlisting}
\stitle{Case 4: Lists in the RDF Model} The RDF model utilizes composite types such as lists, bags, and sequences using an RDF container.\footnote{\url{https://www.w3.org/TR/rdf-schema/\#ch_container}} Case 4 represents an RDF list of three elements. Any transformation approach from RDF graph to property graph should be able to represent the different features of the RDF-star model, such as RDF lists, in the generated property graph. In the property graph, arrays can be the equivalent representation of the RDF list. However, different transformation methods might have different specifications for handling this representation, and that is what we test using this case.
\begin{lstlisting}[
style={st:rdf},
label={lst:CollectionRDF}
]
#Case 4:
@prefix ex: <http://example.org/> .
ex:List1 ex:contents ("one" "two" "three") .
\end{lstlisting}

\stitle{Case 5: Blank Nodes} 
The blank node or the anonymous resource in the RDF model represents a resource for which a URI is unavailable. As per the W3C standards\footnote{\url{https://www.w3.org/TR/2014/REC-rdf11-concepts-20140225/}}, a blank node can only be used as the subject or object of an RDF triple. In Case 5 example, the two RDF triples have the blank node (\_:c) as an object and a subject. Transforming an RDF triples with anonymous resources to a property graph should be tested. 
\begin{lstlisting}[
style={st:rdf},
label={lst:case13}
]
#Case 5:
@prefix ex: <http://example.org/> .
@prefix rdf: <http://www.w3.org/1999/02/22-rdf-syntax-ns> .
ex:bob ex:nationality _:c .
_:c a  ex:Person .
\end{lstlisting}
\stitle{Case 6: Named Graph}
It is possible to assign a name for the RDF graph for access control, partitioning, or to attach metadata to the overall fragment of the RDF graph using the notion of named graphs. By default, the \emph{RDF graph} does not have a name and may not contain any triples, or it contains zero or more named graphs. In the RDF example in Case 6, (Artist) represents an RDF graph. The concept of named graphs in the RDF model has no equivalent in the property graphs.
\begin{lstlisting}[style={st:rdf},
label={lst:ngraphs}
]
#Case 6:
@prefix rdf: <http://www.w3.org/1999/02/22-rdf-syntax-ns> .
@prefix rdfs: <http://www.w3.org/2000/01/rdf-schema> .
@prefix ex: <http://example.org/> .
ex:Graph1 { ex:Monica ex:name "Monica" .      
            ex:Monica ex:homepage ex:Monicahompage .
            ex:Monica ex:hasSkill ex:Management }
ex:Graph2 { ex:Monica rdf:type ex:Person .
            ex:Monica ex:hasSkill ex:Programming }
\end{lstlisting}
\stitle{Case 7: Multi-Labels or Types for RDF Resources} 
The RDF model allows the RDF graph to contain data and schema, such as using \emph{rdf:type} or \emph{a} RDF predicate in the RDF and RDF-star triples. In this example, we evaluate the case when the same RDF resource is associated with multiple RDF types. For instance, in Case 7 example, the RDF resource (ex:alice) is associated with two different RDF types (ex:Artist) and (ex:Author). The example might be challenging for certain property graph representations, especially if the transformation approach does not consider separating the RDF types from the RDF data.  
\begin{lstlisting}[
style={st:rdf},
label={lst:case14}
]
#Case 7:
@prefix ex: <http://example.com/> .
@prefix rdf: <http://www.w3.org/1999/02/22-rdf-syntax-ns> .
ex:alice  a ex:Artist .
ex:alice  a ex:Author .
\end{lstlisting}

\subsection{RDF-star}
\label{App:rdfs}

\stitle{Case 8: Embedded object property statement in subject position} 
As the name indicates and the following listing shows, Case 8 test case features an RDF-star statement where the subject corresponds to an embedded object property statement and the object is a literal: 

\begin{lstlisting}[
style={st:rdf},
label={lst:rdfstarexample}
]
#Case 8:
@prefix ex: <http://example.org/> .
<<ex:alice ex:likes ex:bob>> ex:certainty 0.5 .
\end{lstlisting}


\stitle{Case 9: Embedded datatype property statement in subject position} 
Similar to the previous case we again have an RDF-star statement where the subject corresponds to an embedded statement. In contrast to the Case 8, the embedded statement in Case 9 is a datatype property statement:
\begin{lstlisting}[
style={st:rdf},
label={lst:case4}
]
#Case 9:
@prefix ex: <http://example.org/> .
<<ex:Mark ex:age 28>> ex:certainty 1 .
\end{lstlisting}
\stitle{Case 10: Embedded object property statement in object position} 
Of course, RDF-star statements can also have embedded statements on object position, which is covered in this case. Similar to Case 8, the embedded statement is an \emph{object property} statement. 
\begin{lstlisting}[
style={st:rdf},
label={lst:case5}
]
#Case 10:
@prefix ex: <http://example.org/> .
ex:bobhomepage ex:source <<ex:mainPage ex:writer ex:alice>> .
\end{lstlisting}

\stitle{Case 11: Embedded object property triple as a subject and non-literal object} 
Unlike Cases 8 and 9, where the object is a literal value, the example of case 11.1 represents an RDF-star triple that consists of an embedded triple as a subject (<<ex:mainPage ex:writer ex:alice>>), a predicate (ex:source), and an object as an IRI resource (ex:bobhomepage). 
\begin{lstlisting}[
style={st:rdf},
label={lst:Case6.1}
]
#Case 11.1:
@prefix ex: <http://example.org/> .
<<ex:mainPage ex:writer ex:alice>> ex:source ex:bobhomepage .
\end{lstlisting}
Another variation is in Case 11.2, such that the subject is an embedded triple (<<ex:alice ex:friend ex:bob>>) that has the predicate \url{ex:mentionedBy} and the IRI resource \url{ex:Alex} as an object. Additionally, the object is a subject in a different triple in the same RDF graph.
\begin{lstlisting}[
style={st:rdf},
label={lst:Case6.2}
]
#Case 11.2:
@prefix ex: <http://example.org/> .
<<ex:alice ex:friend ex:bob>> ex:mentionedBy ex:Alex .
  ex:Alex  ex:age    25 . 
\end{lstlisting}

\stitle{Case 12: Embedded object property triple as a subject with \emph{RDF:Type} predicate} Case 12 represents an RDF-star triple that consists of an embedded triple as a subject, a predicate, and a non-literal object. The intuition behind this case is to test the transformation of the RDF-star triples that contain elements of the RDF schema. In Case 12.1, we use \emph{rdf:type} as a predicate of the outer triple:
\begin{lstlisting}[
style={st:rdf},
label={lst:Case7.1}
]
#Case 12.1:
@prefix rdf: <http://www.w3.org/1999/02/22-rdf-syntax-ns#>.
@prefix ex: <http://example.com/> .
@prefix rdfs: <http://www.w3.org/2000/01/rdf-schema#> .
<<ex:mainPage ex:writer ex:alice>> rdf:type ex:bobhomepage .
\end{lstlisting}
In Case 12.2, we use \emph{rdf:type} as a predicate of the embedded triple as follows:
\begin{lstlisting}[
style={st:rdf},
label={lst:Case7.2}
]
#Case 12.2:
@prefix rdf: <http://www.w3.org/1999/02/22-rdf-syntax-ns#> .
@prefix ex: <http://example.org/> .
<<ex:lara rdf:type ex:writer>> ex:owner ex:Journal .

\end{lstlisting}
\stitle{Case 13: Double nested RDF-star triple in the subject position}
This case is considered beyond the basics of the RDF-star model. In Case 13 example, we test the double nesting of RDF-star triple such that an RDF-star triple is embedded in another RDF-star triple. 
\begin{lstlisting}[
style={st:rdf},
label={lst:Case8}
]
#Case 13:
@prefix ex: <http://example.com/> .
<<<<ex:Steve ex:position "CEO">> ex:mentionedBy ex:book>> ex:source ex:journal .
\end{lstlisting}



\stitle{Case 14: Multi-value properties in RDF and RDF-star models}
The RDF model supports multi-value properties, whereas a property graph in many implementations usually just supports mono-value properties~\cite{tomaszuk2019serialization}. RDF multi-value properties triples are the triples where the subject and predicate are the same, but the object is different. In Case 14.1, the two RDF triples have the same subject and predicate, but different objects ("Info\_Page" and "aau\_page"). 
\begin{lstlisting}[
style={st:rdf},
label={lst:multivlsRDF}
]
#Case 14.1:
@prefix ex: <http://example.org/> .
ex:college_page ex:subject "Info_Page" ; 
                ex:subject "aau_page" .
\end{lstlisting}
A natural extension for this case is to evaluate the transformation of multi values for an embedded triple. In Case 14.2, the two RDF-star triples have the same subject, the embedded triple (<<ex:Mary ex:likes ex:Matt>>), and the same predicate (ex:certainty), but different literal values as objects (0.5 and 1).
\begin{lstlisting}[
style={st:rdf},
label={lst:multivlsRDFstar}
]
#Case 14.2:
@prefix ex: <http://example.org/> .
<<ex:Mary ex:likes ex:Matt>> ex:certainty 0.5 .
<<ex:Mary ex:likes ex:Matt>> ex:certainty 1 .
\end{lstlisting}

\stitle{Case 15: Multiple Edge Instances in the Property Graphs} 
The property graph can have multiple edges between the same source and destination nodes. For example, suppose we want to describe a statement that a person has met another person in different times, probably in different occasions. In that case, we can describe it with the property graph by creating multiple edges with different properties between the same pair of nodes (the two people), but we cannot describe this with the RDF graph. In theory, the RDF graph is a set of triples that each triple can occur just once, and it is impossible to have multiple edges of the same type between the same pair of nodes. Similarly, in the RDF-star graph, any embedded triple is unique. Thus, it cannot exist another instance of that triple in the same graph. This case would be challenging to be handled by the transformation approaches for mapping a property graph to an RDF-star graph, not the other way around, which is out of the scope of our research.
Alternatively, we consider a case where the single RDF-star graph contains a set of identical embedded triples in different RDF-star triples and evaluate their transformation to the property graph. In Case 15.1, the embedded triple is used in two different RDF-star triples with different predicates \url{ex: certainty} and \url{ex: source} and objects \url{0.5} and \url{"text"}. 
\begin{lstlisting}[
style={st:rdf},
label={lst:multirdfstar}
]
#Case 15.1:
@prefix ex: <http://example.com/> .
<<ex:Mary ex:likes ex:Matt>> ex:certainty 0.5 .
<<ex:Mary ex:likes ex:Matt>> ex:source "text" .
\end{lstlisting}
In Case 15.2, the embedded triple (ex:Mary ex:likes ex:Matt) exists two times in the single graph: one time as a subject of an RDF-star triple and another as an independent triple.  
\begin{lstlisting}[
style={st:rdf},
label={lst:multiembeded}
]
#Case 15.2:
@prefix ex: <http://example.com/> .
<<ex:Mary ex:likes ex:Matt>> ex:certainty 0.5 .
  ex:Mary ex:likes ex:Matt .
\end{lstlisting}

%% file: experiments.tex
In this section, we present the output of the test cases upon evaluating different transformation approaches—notably, the transformation from RDF and RDF-star to property graphs.

\stitle{Results:} Table~\ref{tab:mapping} presents the output of mapping different RDF and RDF-star graphs to property graphs in the three project. The transformation output for each project is expressed by the formalization for the property graph model in Section~\ref{sec:preliminaries}. The results mainly differ in the structure of the property graph generated based on test cases or whether the test cases are supported or not. In Table~\ref{tab:mapping}, we use \emph{Not supported} to denote that the project does not consider supporting this case in the research or the implementation. We use \emph{Triple is ignored} when the project is claimed to handle the case, but when tested, we found that it skips RDF triples without transforming them to a property graph or producing any exceptional error. \emph{Parsing error} indicates that the project shows an error while reading the input graph.

\clearpage%
\pagenumbering{gobble}
\begingroup 
\setlength{\LTleft}{-4cm}
\begin{landscape}%
\setlength\tabcolsep{2pt}
 \fontsize{7}{11}\selectfont
\input{LongTable} 
\end{landscape}

\endgroup
\clearpage


%% file: LongTable.tex
\begin{longtable}[th!]{cc|c|c|c|} 
 \\ \hline
 \multicolumn{1}{|c|} {} & \textbf{Triples} &  \textbf{Neosemantics~\cite{neosemantics}}  & \textbf{RDF-Star Tools~\cite{RDFReconciliation}} &  \textbf{RDF2PG~\cite{angles2020mapping}}  \\ \hline
\endhead

\multicolumn{5}{|c|} {\textbf{RDF}}
\\ \hline 

 \multicolumn{1}{|p{1.5cm}|}{
 \centering
 Case 1

 } 

 & 

\multicolumn{1}{l|}{ 
\begin{tabular}[l]{@{}l@{}} 

@prefix ex: <http://example.com/> .
\\
\\ex:alice ex:meets ex:bob .
\end{tabular}
 } 

& \multicolumn{1}{l|}{ 
\begin{tabular}[l]{@{}l@{}}  
    $N=$\{$n_{1}, n_{2}$\}
\\  $E= $\{$e_{1}$\}
 \\ $\forall i \in$ \{$1,2$\}; $lbl(n_{i})=$ \{"Resource"\} 
\\  $edge(e_{1})$= ($n_{1}, n_{2}$) 
  \\$\sigma(n_{1})$= \{("url",\\ "http://example.com/alice")\}
   \\$\sigma(n_{2})$= \{("url",\\ "http://example.com/bob")\}
 \\$\sigma(e_{1})$= \{("type", \\ "$localName$(http://example.com/meets)")\}
\end{tabular}
 }  &  
 \multicolumn{1}{l|} {\begin{tabular}[l]{@{}l@{}}  
$N=$\{$n_{1}, n_{2}$\}
\\  $E=$\{$e_{1}$\}
\\ $edge(e_{1})$= ($n_{1}, n_{2}$)
\\ $lbl(e_{1})=$ \{"http://example.com/meets"\}
   \\$\sigma(n_{1})$= \{("Kind","IRI"),\\ ("IRI", "http://example.com/alice")\}
   \\$\sigma(n_{2})$= \{("Kind","IRI"),\\ ("IRI", "http://example.com/bob")\}
\end{tabular}}

 & 

\multicolumn{1}{l|}{ 
\begin{tabular}[l]{@{}l@{}}  
 \\  $N=$\{$n_{1}, n_{2}$\}
\\  $E= $\{$e_{1}$\}
\\ $edge(e1)$ =($n_{1}, n_{2}$)
\\ $\forall i \in$ \{$1,2$\}; $lbl(n_{i})=$ \{"Resource"\} 
\\ $lbl(e1)=$ \{"ObjectProperty"\}
   \\$\sigma(n_{1})$= \{("iri",\\ "http://example.com/alice")\}
   \\$\sigma(n_{2})$= \{("iri",\\ "http://example.com/bob")\}
   
  \\ $\sigma(e1)$= \{("type",
 \\"http://example.com/meets")\}

\end{tabular}
 }
 \\ \hline %
 \multicolumn{1}{|p{1.5cm}|} {  \centering Case 2.1 }  &  

\multicolumn{1}{l|}{ 
\begin{tabular}[l]{@{}l@{}} 

@prefix ex: <http://example.com/> .
\\
\\ ex:Sam ex:mentor ex:Lee.
\\ ex:mentor \\ rdfs:label "project supervisor".
\\ ex:mentor \\ex:name "mentor's name".

\end{tabular}
 }

& \multicolumn{1}{l|}{ 

\begin{tabular}[l]{@{}l@{}}  
    $N=$\{ $ n_{1}, n_{2},n_{3} $\}
\\  $E= $\{$e_{1}$\}
\\  $edge(e_{1}) =$ ($n_{1}, n_{2}$)
   \\$\forall i \in$ \{$1,2,3$\}; $lbl(n_{i})=$ \{"Resource"\} 
  \\$\sigma(n_{1})$=  \{("url",\\ "http://example.com/Sam")\}
   \\$\sigma(n_{2})$=  \{("url", \\"http://example.com/Lee")\}
     \\$\sigma(n_{3})$=  \{("url", \\"http://example.com/mentor")\}
      \\ $\sigma(n_{3})$= \{("name", "mentor's name"),\\ ("label","project supervisor")\}
 \\$\sigma(e_{1})$= \{("type", \\"$localName$("http://example.com/mentor)")\}

\end{tabular}
}  & 
 \multicolumn{1}{l|} {\begin{tabular}[l]{@{}l@{}}  
\\
$N= $\{$ n_{1}, n_{2}, n_{3}, n_{4}, n_{5}$ \}
\\  $E=$\{$e_{1}, e_{2}, e_{3}$\}
\\  $edge(e_{1})$= ($n_{3}, n_{5}$)
\\  $edge(e_{2})$= ($n_{1}, n_{2}$)
\\  $edge(e_{3})$= ($ n_{3}, n_{4}$)
\\ $lbl(e_{1})=$ \{"http://example.com/name"\}
\\ $lbl(e_{2})=$ \{"http://example.com/mentor"\}
\\ $lbl(e_{3})=$  \{"http://www.w3.org\\/2000/01/rdf-schema\#label"\}
 
 \\$\sigma(n_{1})$= \{("Kind","IRI"),\\ ("IRI", "http://example.com/Sam")\}
 
  \\$\sigma(n_{2})$= \{("Kind","IRI"),\\ ("IRI", "http://example.com/Lee")\}
 
  \\$\sigma(n_{3})$= \{("Kind","IRI"),\\ ("IRI", "http://example.com/mentor")\}
 
 \\ $\sigma(n_{4})$= \{("Kind","Literal"), 
  \\ ("Literal","project supervisor")\}
 \\ $\sigma(n_{5})$= \{("Kind","Literal"), 
  \\ ("Literal","mentor's name")\} 
\end{tabular}
}

& 

\multicolumn{1}{l|}{ 
\begin{tabular}[l]{@{}l@{}}  
 \\  $N=$\{$n_{1}, n_{2}, n_{3}, n_{4}, n_{5}$\}
\\  $E= $\{$e_{1}, e_{2}, e_{3}$\}
 \\ $edge(e_{1})$= ($n_{1}, n_{2}$)
 \\ $edge(e_{2})$= ($n_{3}, n_{4}$)
  \\ $edge(e_{3})$= ($n_{3}, n_{5}$)

 \\  $\forall i \in$ \{$1,2,3$\}; $lbl(n_{i})=$ \{"Resource"\}
 \\ $\forall i \in$ \{$4,5$\}; $lbl(n_{i}) =$ \{"Literal"\}
 
 \\ $lbl(e_{1}) =$ \{"ObjectProperty"\}
  \\  $\forall i \in$ \{$2,3$\}; $lbl(e_{i}) =$\{"DatatypeProperty"\}
\\$\sigma(n_{1})$=  \{("iri",\\ "http://example.com/Sam")\}
   \\$\sigma(n_{2})$=  \{("iri", \\"http://example.com/Lee")\}
   \\$\sigma(n_{3})$=  \{("iri", \\"http://example.com/mentor")\}  
\\  $\sigma(n_{4})$= \{("value",
\\ "project supervisor")\}
  \\   $\sigma(n_{5})$= 
  \\ \{ ("value", "mentor's name")\}
   \\ $\sigma(e_{1})$= \{("type",\\ "http://example.com/mentor")\}
 \\ $\sigma(e_{2})$= \{("type",  "http://www.w3.org/\\ 2000/01/rdf-schema\#label")\}
  \\ $\sigma(e_{3})$= \{("type",\\ "http://example.com/name")\}

\end{tabular}
 }
  \\ \hline
 \multicolumn{1}{|p{1.5cm}|} {  \centering Case 2.2  } 
 & \multicolumn{1}{l|}{ 
\begin{tabular}[l]{@{}l@{}} 

@prefix ex: <http://example.com/> .
\\
\\ ex:Martin ex:mentorJoe ex:Joe.
\\ ex:mentorJoe ex:alias ex:teacher . 						
\end{tabular}
 }  & 
 
 \multicolumn{1}{l|}{ 
 \begin{tabular}[l]{@{}l@{}}  
    $N=$\{$n_{1}, n_{2}, n_{3}, n_{4}$\}
\\  $E= $\{$e_{1}, e_{2}$\}
\\$edge(e_{1})$= ($n_{1}, n_{2}$)
\\ $edge(e_{2})$= ($n_{3}, n_{4}$)
  \\ $\forall i \in$ \{$1, 2, 3, 4$\}; $lbl(n_{i})$= \{"Resource"\}
  
  \\$\sigma(n_{1})$= \{("url",\\ "http://example.com/Martin")\}
   \\$\sigma(n_{2})$=  \{("url",\\ "http://example.com/Joe")\}
   
    \\$\sigma(n_{3})$=  \{("url",\\ "http://example.com/mentorJoe")\}
     \\$\sigma(n_{4})$=  \{("url",\\ "http://example.com/teacher")\}
  \\$\sigma(e_{1})$= \{("type",
  \\"$localName$(http://example.com/mentorJoe)")\}
  \\$\sigma(e_{2})$= \{("type",
  \\ "$localName$(http://example.com/alias)")\}
\end{tabular}
 } 

 &  \multicolumn{1}{l|} {\begin{tabular}[l]{@{}l@{}}  
$N= $\{$ n_{1}, n_{2}, n_{3}, n_{4}$\}
\\  $E=$\{$e_{1}, e_{2}$\}
\\$edge(e_{1})$= ($n_{3}, n_{4}$)
\\$edge(e_{2})$= ($n_{1}, n_{2}$)
 \\$lbl(e_{1})=$\{"http://example.com/alias"\}
\\ $lbl(e_{2})=$\{"http://example.com/mentorJoe"\}
  \\ $\sigma(n_{1})$= \{("Kind","IRI"), ("IRI", \\"http://example.com/Martin")\}

   \\ $\sigma(n_{2})$= \{("Kind","IRI"), ("IRI", \\"http://example.com/Joe")\}
    \\ $\sigma(n_{3})$= \{("Kind","IRI"), ("IRI", \\"http://example.com/mentorJoe")\}
   \\ $\sigma(n_{4})$= \{("Kind","IRI"), ("IRI", \\"http://example.com/Teacher")\}

\end{tabular}
}

 & 
 
  \multicolumn{1}{l|}{ 
\begin{tabular}[l]{@{}l@{}}  
 \\  $N=$\{$n_{1}, n_{2}, n_{3}, n_{4}$\}
\\  $E= $\{$e_{1}, e_{2}$\}
 \\ $edge(e_{1})$= ($n_{1}, n_{2}$)
  \\ $edge(e_{2})$= ($n_{3}, n_{4}$)
  \\ $\forall i \in$ \{$1,2,3,4$\}; $lbl(n_{i})=$ \{"Resource"\}
   \\  $\forall i \in$ \{$1,2$\}; $lbl(e_{i}) =$ \{"ObjectProperty"\}
   
  \\  $\sigma(n_{1})$= \{("iri",
 \\ "http://example.com/Martin")\}
 
   \\  $\sigma(n_{2})$= \{("iri",
 \\ "http://example.com/Joe")\}
    \\  $\sigma(n_{3})$= \{("iri",
 \\ "http://example.com/mentorJoe")\}
    \\  $\sigma(n_{4})$= \{("iri",
 \\ "http://example.com/Teacher")\}
 \\ $\sigma(e_{1})$= \{("type",
 \\ "http://example.com/mentorJoe")\}
  \\ $\sigma(e_{1})$= \{("type","http://example.com/alias")\}
\end{tabular}
 }
  \\ \hline

 \multicolumn{1}{|p{1.5cm}|} { \centering Case 2.3}  & 

\multicolumn{1}{l|}{ 
\begin{tabular}[l]{@{}l@{}} 

@prefix rdfs: \\ <http://www.w3.org/2000/01/
\\ rdf-schema\#> .
\\@prefix ex: <http://example.com/> .
\\
\\ex:Jan ex:supervise ex:Leo.
\\ ex:supervise \\ rdfs:subPropertyOf ex:administer .					
\end{tabular}
 }
  & 
  
  \multicolumn{1}{l|}{ 

 \begin{tabular}[l]{@{}l@{}}  
    $N=$\{$n_{1}, n_{2}, n_{3}, n_{4}$\}
\\  $E= $\{$e_{1}, e_{2}$\}
 \\ $edge(e_{1})$= ($n_{1}, n_{2}$)
  \\$edge(e_{2})$= ($n_{3}, n_{4}$)
   \\ $\forall i \in$ \{$1, 2, 3, 4$\}; $lbl(n_{i})$ = \{"Resource"\} 
  \\$\sigma(n_{1})$= \{("url","http://example.com/Jan")\}
   \\$\sigma(n_{2})$= \{("url","http://example.com/Leo")\} 
    \\$\sigma(n_{3})$= \{("url","http://example.com/supervise")\}
    
     \\$\sigma(n_{4})$= \{("url","http://example.com/administer")\}

       \\$\sigma(e_{1})$= \{("type",\\ "$localName$(http://example.com/supervise)")\}
    \\$\sigma(e_{2})$= \{("type",\\ "$localName$(http://www.w3.org/\\ 2000/01/rdf-schema\#subPropertyOf)")\}
\end{tabular}
 } 
  & 
   \multicolumn{1}{l|} {\begin{tabular}[l]{@{}l@{}}  
$N= $\{$ n_{1}, n_{2}, n_{3}, n_{4}$\}
\\  $E=$\{$e_{1}, e_{2}$\}
 \\ $edge(e_{1})$= ($n_{3}, n_{4}$)
  \\ $edge(e_{2})$= ($n_{1}, n_{2}$)
 \\$lbl(e_{1})=$ "http://www.w3.org/\\ 2000/01/rdf-schema\#subPropertyOf"
 \\ $lbl(e_{2})=$ "http://example.com/supervise"
 \\ $\sigma(n_{1})$= \{("Kind","IRI"),\\ ("IRI","http://example.com/Jan")\}
  \\ $\sigma(n_{2})$= \{("Kind","IRI"),\\ ("IRI","http://example.com/Leo")\}
   \\ $\sigma(n_{3})$= \{("Kind","IRI"),\\ ("IRI","http://example.com/supervise")\}
   \\ $\sigma(n_{4})$= \{("Kind","IRI"),\\ ("IRI","http://example.com/administer" )\}

\end{tabular}
} 
   & 
    \multicolumn{1}{l|}{ 
\begin{tabular}[l]{@{}l@{}}  
 \\  $N=$\{$n_{1}, n_{2}, n_{3}, n_{4}$\}
\\  $E= $\{$e_{1}, e_{2}$\}
 \\ $edge(e_{1})$= ($n_{1}, n_{2}$)
  \\$edge(e_{2})$= ($n_{3}, n_{4}$)
 
    \\ $\forall i \in$ \{$1,2,3,4$\}; $lbl(n_{i})=$ \{"Resource"\}  
    \\  $\forall i \in $\{$1,2$\}; $lbl(e_{i}) =$ \{"ObjectProperty"\}
 \\ $\sigma(n_{1})$= \{("iri","http://example.com/Jan")\}
  \\ $\sigma(n_{2})$= \{("iri","http://example.com/Leo")\} 
   \\ $\sigma(n_{3})$= \{("iri","http://example.com/supervise")\} 
     \\ $\sigma(n_{4})$= \{("iri","http://example.com/administer")\} 
 \\ $\sigma(e_{1})$= \{("type","http://example.com/supervise")\}
  \\ $\sigma(e_{2})$= \{("type",
  "http://www.w3.org/\\ 2000/01/rdf-schema\#subPropertyOf")\}
 
\end{tabular}
 }
 \\ \hline
 
 \multicolumn{1}{|p{1.5cm}|} { \centering Case 2.4} & 
 
 \multicolumn{1}{l|}{ 
\begin{tabular}[l]{@{}l@{}} 
\\@prefix ex: <http://example.com/> .
\\@prefix rdf: <http://www.w3.org/\\1999/ 02/22-rdf-syntax-ns\#>.
\\
\\ ex:Tom ex:friend ex:Chris.
\\ ex:friend rdf:type ex:relation .				
\end{tabular}
 }
 
 & 
 
 \multicolumn{1}{l|}{ 
 
 \begin{tabular}[l]{@{}l@{}}  
       $N=$\{$n_{1}, n_{2}, n_{3}$\}
\\  $E= $\{$e_{1}$\}
 \\ $edge(e_{1})$= ($n_{1}, n_{2}$)
    \\ $\forall i \in$ \{$1, 2$\}; $lbl(n_{i})$= \{"Resource"\} 
  \\$lbl(n_{3}) =$ \{"Resource", "relation"\}
   \\$\sigma(n_{1})$= \{("url","http://example.com/Tom")\}
    \\$\sigma(n_{2})$= \{("url","http://example.com/Chris")\}
     \\$\sigma(n_{3})$= \{("url","http://example.com/friend")\}
 \\$\sigma(e_{2})$=  \{("type",\\ "$localName$(http://example.com/friend)")\}
\end{tabular}
 } 
 
 & 
 \multicolumn{1}{l|} {\begin{tabular}[l]{@{}l@{}}  
$N= $\{$ n_{1}, n_{2}, n_{3}, n_{4}$\}
\\  $E=$\{$e_{1}, e_{2}$\}
\\ $edge(e_{1})$= ($n_{3}, n_{4}$)
 \\$lbl(e_{1})=$ \{"http://www.w3.org/\\ 1999/02/22-rdf-syntax-ns\#type"\}
 \\ $edge(e_{2})$= ($n_{1}, n_{2}$)
\\ $lbl(e_{2})=$ \{"http://example.com/friend"\}
 \\ $\sigma(n_{1})$= \{ ("Kind","IRI"),\\
 ("IRI", "http://example.com/Tom")\}
 \\ $\sigma(n_{2})$= \{ ("Kind","IRI"),\\
 ("IRI", "http://example.com/Chris")\}
 \\ $\sigma(n_{3})$= \{ ("Kind","IRI"),\\
 ("IRI", "http://example.com/friend")\}
  \\ $\sigma(n_{4})$= \{ ("Kind","IRI"),\\
 ("IRI", "http://example.com/relation")\}
 
\end{tabular}
}
 
 & 
 
 \multicolumn{1}{l|}{ 
\begin{tabular}[l]{@{}l@{}}  
 \\  $N=$\{$n_{1}, n_{2}, n_{3}, n_{4}$\}
\\  $E= $\{$e_{1}, e_{2}$\}
 \\ $edge(e_{1})$= \{$n_{1}, n_{2}$\}
\\ $edge(e_{2})$= \{$n_{3}, n_{4}$\}
 \\ $\forall i \in$ \{$1,2$\}; $lbl(_{i})=$ \{"ObjectProperty"\} 
 \\$\forall i \in$ \{$1,2,3,4$\}; $lbl(n_{i}) =$ \{"Resource"\} 
  \\ $\sigma(n_{1})$= \{("iri", "http://example.com/Tom")\}
   \\ $\sigma(n_{2})$= \{("iri", "http://example.com/Chris")\}
    \\ $\sigma(n_{3})$= \{("iri", "http://example.com/friend")\}
     \\ $\sigma(n_{4})$= \{("iri", "http://example.com/relation")\}
 \\ $\sigma(e_{1})$= \{("type", "http://example.com/friend")\}
  \\ $\sigma(e_{2})$= \{("type",
  "http://www.w3.org/\\ 1999/02/22-rdf-syntax-ns\#type")\}
\end{tabular}
 }

  \\ \hline

   \multicolumn{1}{|p{1.5cm}|} { \centering Case 3.1} &  
  \multicolumn{1}{l|}{ 
\begin{tabular}[l]{@{}l@{}} 
\\@prefix ex: <http://example.com/> .
\\
\\ ex:book  ex:publish\_date\\ "1963-03-22"$^\wedge$$^\wedge$ xsd:date.
\\ ex:book  ex:pages "100"$^\wedge  $$^\wedge$xsd:integer.
\\  ex:book ex:cover 20 .
\\ ex:book ex:index "55".
\end{tabular}
 }

&
\multicolumn{1}{l|}{ 
 \begin{tabular}[l]{@{}l@{}} 
 \\  $N=$\{$n_{1}$\}
 \\$E= \emptyset$
  \\$lbl(n_{1}) =$ \{"Resource"\}
  \\$\sigma$($n_{1}$) = \{("index", 55),
  \\("url","http://example.com/book")
  \\("pages", 100), ("cover", 20),
  \\("publish\_date", "1963-03-22")\}
 
\end{tabular}
}

&
 \multicolumn{1}{l|} {\begin{tabular}[l]{@{}l@{}}  
$N= $\{$ n_{1}, n_{2}, n_{3}, n_{4}, n_{5}$\}
\\  $E=$\{$e_{1}, e_{2}, e_{3}, e_{4}$\}
 \\ $edge(e_{1})$= ($n_{1}, n_{5}$)
  \\ $edge(e_{2})$= ($n_{1}, n_{3}$)
   \\ $edge(e_{3})$= ($n_{1}, n_{4}$)
    \\ $edge(e_{4})$= ($n_{1}, n_{2}$)
 \\$lbl(e_{1})=$ \{"http://example.com/index"\}
 \\$lbl(e_{2})=$ \{"http://example.com/pages"\}
 \\$lbl(e_{3})=$ \{"http://example.com/cover"\}
 \\$lbl(e_{4})=$ \{"http://example.com/publish\_date"\}
 
 \\ $\sigma(n_{1})$= \{("Kind","IRI"), \\
 ("IRI", "http://example.com/book") \}
   \\ $\sigma(n_{2})$= \{("Kind","Literal"),\\
  ("Literal", 1963-03-22),\\("Datatype","http://www.w3.org/\\2001/XMLSchema\#date")\}
   \\ $\sigma(n_{3})$= \{("Kind","Literal"),
  ("Literal", 100),\\("Datatype","http://www.w3.org\\/2001/XMLSchema\#integer")\}
     \\ $\sigma(n_{4})$= \{("Kind","Literal"),
  ("Literal", 20),\\("Datatype","http://www.w3.org\\/2001/XMLSchema\#integer")\}
  \\ $\sigma(n_{5})$= \{("Kind","Literal"),
  ("Literal", 55),\\("Datatype", "http://www.w3.org\\/2001/XMLSchema\#string")\}
\end{tabular}
}

& 
\multicolumn{1}{l|}{ 
\begin{tabular}[l]{@{}l@{}}  
 \\  $N=$\{$n_{1}, n_{2}, n_{3}, n_{4}, n_{5}$\}
\\  $E= $\{$e_{1}, e_{2}, e_{3}, e_{4}$\}
 \\ $edge(e_{1})$= ($n_{1}, n_{2}$)
 \\ $edge(e_{2})$= ($n_{1}, n_{3}$)
  \\ $edge(e_{3})$= ($n_{1}, n_{4}$)
   \\ $edge(e_{4})$= ($n_{1}, n_{5}$)
  \\ $lbl(n_{1}) =$ \{"Resource"\} 
\\ $\forall i \in$ \{$2,3,4$\}; $lbl(n_{2,3,4}) =$ \{"Literal"\}
\\  $\forall i \in$ \{$1,2,3,4$\}; $lbl(e_{i})$= \{"DatatypeProperty"\}

 \\$\sigma(n_{1})$= \{ ("iri", "http://example.com/book")\}
   \\$\sigma(n_{2})$= \{ ("value", "1963-03-22"),
   ("type",\\ "http://www.w3.org/2001/XMLSchema\#date")\}
     \\$\sigma(n_{3})$= \{ ("value", "100"),
    ("type", \\ "http://www.w3.org/2001/XMLSchema\#integer")\} 
        \\$\sigma(n_{4})$= \{ ("value", "20"),
   ("type", \\ "http://www.w3.org/\\2001/XMLSchema\#integer")\}
      \\$\sigma(n_{5})$= \{ ("value", "55"),
    ("type", \\ "http://www.w3.org/\\2001/XMLSchema\#string")\} 
   
          \\$\sigma(e_{1})$= \{ ("type", "http://example.com/publish\_date")
           \\$\sigma(e_{2})$= \{ ("type", "http://example.com/pages")
           \\$\sigma(e_{3})$= \{ ("type", "http://example.com/cover")
            \\$\sigma(e_{4})$= \{ ("type", "http://example.com/index")
\end{tabular}
 }
 
 \\ \hline
  \multicolumn{1}{|p{1.5cm}|} { \centering Case 3.2}  & 
    \multicolumn{1}{l|}{ 
\begin{tabular}[l]{@{}l@{}} 
\\@prefix ex: <http://example.com/> .
\\
\\ex:book  ex:Englishtitle "Book"@en.
\\ex:book  ex:title "Bog"@da.
\end{tabular}
 }

  &\multicolumn{1}{l|}{ 
 \begin{tabular}[l]{@{}l@{}} 
 \\  $N=$\{$n_{1}$\}
 \\$E= \emptyset$
  \\$lbl(n_{1}) =$ \{"Resource"\}
  \\$\sigma$($n_{1}$) =("Englishtitle",
  "Book@en"),\\ \{("url","http://example.com/book"),
  \\("title", "Bog@da")\}
\end{tabular}
}
 & 
 
    \multicolumn{1}{l|} {\begin{tabular}[l]{@{}l@{}}  
$N= $\{$ n_{1}, n_{2}, n_{3}$\}
\\  $E=$\{$e_{1}, e_{2}$\}
 \\ $edge(e_{1})$= ($n_{1}, n_{2}$)
  \\ $edge(e_{2})$= ($n_{1}, n_{3}$)
 \\$lbl(e_{1})=$ \{"http://example.com/Englishtitle"\}
  \\$lbl(e_{2})=$\{"http://example.com/title"\}
 \\ $\sigma(n_{1})$= \{("Kind","IRI"), 
 \\("IRI, "http://example.com/book")\}
  \\ $\sigma(n_{2})$= \{("Kind","Literal"),
 \\("Literal","Book"), ("Language","en")\}
   \\ $\sigma(n_{3})$= \{("Kind","Literal"),
 \\ ("Literal","Bog"), ("Language","da")\}

\end{tabular}
}
 &
 
  \multicolumn{1}{l|}{ 
\begin{tabular}[l]{@{}l@{}}  
 \\  $N=$\{$n_{1}, n_{2}, n_{3}$\}
\\  $E= $\{$e_{1}, e_{2}$\}
 \\ $edge(e_{1})$= ($n_{1}, n_{2}$)
  \\ $edge(e_{1})$= ($n_{1}, n_{3}$)
\\ $\forall i \in$ \{$2,3$\}; $lbl(n_{i}) =$ \{"Literal"\}
 \\ $\forall i \in$ \{$1,2$\};  $lbl(e_{i}) =$ \{"DatatypeProperty"\}

\\ $\sigma(n_{1})$= \{("iri", "http://example.com/book")\}
   \\$\sigma(n_{2})$= \{("value", "Book"),
   \\ ("type", "http://www.w3.org/1999/02\\/22-rdf-syntax-ns\#langString")\}
     \\$\sigma(n_{3})$= \{("value", "Bog"),
   \\ ("type", "http://www.w3.org/1999/02\\/22-rdf-syntax-ns\#langString")\} 

   \\ $\sigma(e_{1})$= \{("type", "http://example.com/Englishtitle")
      \\ $\sigma(e_{2})$= \{("type", "http://example.com/title")
\end{tabular}
 }
 
  \\ \hline
  
   \multicolumn{1}{|p{1.5cm}|} { \centering Case 4}  & 
 
 \multicolumn{1}{l|}{ 
\begin{tabular}[l]{@{}l@{}} 

@prefix ex: <http://example.com/> .
\\
\\ ex:List1 ex:contents \\ ("one" "two" "three").		
\end{tabular}
 } 
 
 & \multicolumn{1}{l|}{ 
 \begin{tabular}[l]{@{}l@{}} 
  $N=$\{$n_{1},n_{2},n_{3},n_{4}, n_{5}$\}
  \\$E= {e_{1}, e_{2}, e_{3},e_{4}}$
  \\ $edge(e_{1})$= ($n_{1}, n_{2}$)
    \\ $edge(e_{2})$= ($n_{2}, n_{3}$)
\\ $edge(e_{3})$= ($n_{3}, n_{4}$)
\\ $edge(e_{4})$= ($n_{3}, n_{5}$)
\\ $\forall i \in$ \{$1,2,3,4,5$\}; $lbl(n_{i})=$ \{"Resource"\}
\\$\sigma(n_{1})$ = \{("url","http://example.com/List1")\}
  \\$\sigma(n_{2})$ = \{("url","bnode://node1g10t4un5x1"), \\("first", "one")\}
   \\$\sigma(n_{3})$ = \{("url","bnode://node1g10t4un5x2"),\\("first", "two")\}
    \\$\sigma(n_{4})$ = \{("url", "bnode://node1g10t4un5x3"),\\("first", "three")\}

 \\$\sigma(n_{5})$ = \{("url", "http://www.w3.org/1999/02/\\22-rdf-syntax-ns\#nil")\}

  \\$\sigma(e_{1})$ = \{("type",\\"$localName$(http://example.com/contents)")\}
  \\  $\forall i \in$ \{$2,3,4$\};$\sigma(e_{i})$ = \{("type", \\"$localName$(http://www.w3.org/1999/02\\/22-rdf-syntax-ns\#rest")")\}
\end{tabular}
}
&
   \multicolumn{1}{l|} {\begin{tabular}[l]{@{}l@{}}  
\\ $N= $\{$ n_{1}, n_{2}, n_{3}, n_{4}, n_{5}, n_{6},n_{7},n_{8}$\}
\\  $E=$\{$e_{1}, e_{2},e_{3},e_{4},e_{5},e_{6},e_{7}$\}
\\ $edge(e_{1})$= ($n_{1}, n_{2}$)
\\ $edge(e_{2})$= ($n_{8}, n_{1}$)
\\ $edge(e_{3})$= ($n_{3}, n_{5}$)
\\ $edge(e_{4})$= ($n_{1}, n_{3}$)
\\ $edge(e_{5})$= ($n_{5}, n_{6}$)
\\ $edge(e_{6})$= ($n_{3}, n_{4}$)
\\ $edge(e_{7})$= ($n_{5}, n_{7}$)
\\$lbl(e_{2})=$ \{"http://example.com/contents"\}
 \\ $\forall i \in$ \{$1,5,6$\}; $lbl_{e}(e_{i})=$\{"http://www.w3.org/\\1999/02/22-rdf-syntax-ns\#first"\}
 \\ $\forall i \in$ \{$3,4,7$\}; $lbl_{e}(e_{i})=$ \{"http://www.w3.org/\\1999/02/22-rdf-syntax-ns\#rest"\}
 \\ $\sigma(n_{7})$= \{("Kind","IRI"), ("IRI",\\
 "http://www.w3.org/\\1999/02/22-rdf-syntax-ns\#nil")\}
 \\ $\sigma(n_{8})$= \{("Kind","IRI"),\\ ("IRI","http://example.com/List1")\}
  \\ $\forall i \in$ \{$1,3,5$\}; $\sigma(n_{i})$= \{("Kind","Blank")\}
  \\ $\sigma(n_{2})$= \{("Kind","Literal"),
 \\("Literal","One")\}
   \\ $\sigma(n_{4})$= \{("Kind","Literal"),
 \\("Literal","two")\}
    \\ $\sigma(n_{6})$= \{("Kind","Literal"),
 \\("Literal","three")\}

\end{tabular}
}
&
    \multicolumn{1}{l|}{ 
\begin{tabular}[l]{@{}l@{}}  
 \\  $N=$\{$n_{1}, n_{2}, n_{3}, n_{4}, n_{5}, n_{6},n_{7},n_{8}$\}
\\  $E= $\{$e_{1}, e_{2},e_{3},e_{4},e_{5},e_{6},e_{7}$\}
\\ $edge(e_{1})$= ($n_{1}, n_{2}$)
\\ $edge(e_{2})$= ($n_{1}, n_{3}$)
\\ $edge(e_{3})$= ($n_{3}, n_{4}$)
\\ $edge(e_{4})$= ($n_{3}, n_{4}$)
\\ $edge(e_{5})$= ($n_{5}, n_{6}$)
\\ $edge(e_{6})$= ($n_{5}, n_{7}$)
\\ $edge(e_{7})$= ($n_{8}, n_{1}$)
\\ $\forall i \in$ \{$7,8$\}; $lbl(n_{i}) =$ \{"Resource"\}
\\ $\forall i \in$ \{$1,3,5$\}; $lbl(n_{i})=$ \{"BlankNode"\}
\\  $\forall i \in$ \{$2,4,6$\}; $lbl(n_{i})=$ \{"Literal"\}
\\ $\forall i \in$ \{$1,3,5$\}; $lbl(e_{i})=$ \{"DatatypeProperty"\}
\\ $\forall i \in$ \{$2,4,6,7$\}; $lbl(e_{i})=$ \{"ObjectProperty"\}
\\ $\sigma(n_{7})$=\{("iri","http://www.w3.org/1999/\\02/22-rdf-syntax-ns\#nil")\}
\\ $\sigma(n_{8})$= \{("iri","http://example.com/List1")\}
 \\ $\sigma(e_{1,3,5})$=\{("type","http://www.w3.org/1999\\/02/22-rdf-syntax-ns\#first)")\}
  \\ $\sigma(e_{2,4,6})$=\{("type","http://www.w3.org/1999\\/02/22-rdf-syntax-ns\#rest")\}
    \\ $\sigma(e_{7})$=\{("type","http://example.com/contents")\}
\end{tabular}
 }
 \\ \hline
 
  \multicolumn{1}{|p{1.5cm}|} { \centering Case 5} & 
  
  \multicolumn{1}{l|}{ 
\begin{tabular}[l]{@{}l@{}} 

@prefix ex: <http://example.com/> .
\\ ex:bob ex:nationality \_:c .
\\ \_:c a ex:Person.
\end{tabular}
 } 
  &
  \multicolumn{1}{l|}{ 
 \begin{tabular}[l]{@{}l@{}} 
  $N=$\{$n_{1}, n_{2}$\}
 \\$E= {e_{1}}$
  \\ $edge(e_{1})$= ($n_{1}, n_{2}$)
    \\$lbl_{n}(n_{1})=$ \{"Resource"\}
  \\$lbl_{n}(n_{2})=$ \{"Resource","Person"\}
   \\ $\sigma(n_{1})$= \{("url","http://example.com/bob")\}
   \\ $\sigma(n_{2})$= \{("url","bnode://\\genid-e8437ee0b16549e1a53eef561250e4af-c")\}
   \\ $\sigma(e_{1})$ = \{("type",\\ "$localName$(http://example.com/nationality)")\}
  
\end{tabular}
}
&

 \multicolumn{1}{l|} {\begin{tabular}[l]{@{}l@{}}  
\\$N= $\{$ n_{1}, n_{2}, n_{3}$\}
\\  $E=$\{$e_{1}, e_{2}$\}
 \\ $edge(e_{1})$= ($n_{2}, n_{3}$)
  \\ $edge(e_{2})$= ($n_{1}, n_{2}$)
 \\$lbl(e_{1})=$\{"http://www.w3.org/1999/02/\\22-rdf-syntax-ns\#type"\}
 \\$lbl(e_{2})=$ \{"http://example.com/nationality"\}
 \\ $\sigma(n_{1})$= \{("Kind","IRI"), 
\\ ("IRI", "http://example.com/bob")\}
  \\ $\sigma(n_{2})$= \{("Kind","Blank"),
 \\("Blank","c")\}
  \\ $\sigma(n_{3})$= \{("Kind","IRI"), 
 \\("IRI", "http://example.com/Person")\}
  
\end{tabular}
} & 
    \multicolumn{1}{l|}{ 
\begin{tabular}[l]{@{}l@{}}  
   $N=$\{$n_{1}, n_{2}, n_{3}$\}
\\  $E= $\{$e_{1}, e_{2}$\}
 \\ $edge(e_{1})$= ($n_{1}, n_{2}$)
 \\ $edge(e_{2})$= ($n_{1}, n_{3}$)
 \\ $\forall i \in$ \{$1,3$\}; $lbl(n_{i})=$ \{"Resource"\}
\\ $lbl(n_{2})=$ \{"BlankNode"\}
 \\ $\forall i \in$ \{$1,2$\}; $lbl(e_{i})=$ \{"ObjectProperty"\}
 
 \\ $\sigma(n_{1})$= \{("iri","http://example.com/bob")\}
 
  \\ $\sigma(n_{3})$= \{("iri","http://example.com/Person")\}
 
\\ $\sigma(e_{1})$= \{("type","http://example.com/nationality")\}
 \\ $\sigma(e_{2})$= \{("type","http://www.w3.org/1999/02/\\22-rdf-syntax-ns\#type")\}
\end{tabular}
 }
 \\ \hline
 
  \multicolumn{1}{|p{1.5cm}|}{ \centering Case 6} & 
 \multicolumn{1}{l|}{ 
\begin{tabular}[l]{@{}l@{}} 

@prefix ex: <http://example.com/> .
\\
ex:Graph1 \\ \{ ex:Monica ex:name "Monica" .     
          \\ ex:Monica ex:homepage \\ <http://www.Monicahompage.org> .
          \\ ex:Monica \\ ex:hasSkill ex:Management \}
           
\\ex:Graph2 \\ \{ ex:Monica rdf:type ex:Person .
          \\ ex:Monica \\ ex:hasSkill ex:Programming \}.		
\end{tabular}
 } 
 & 
 \multicolumn{1}{l|}{ 
 \begin{tabular}[l]{@{}l@{}} 
\\ $N=$\{$n_{1}, n_{2}, n_{3}, n_{4}$\}
 \\$E= {e_{1}, e_{2}, e_{3}}$
 \\ $edge(e_{1})$= ($n_{1}, n_{2}$)
  \\ $edge(e_{2})$= ($n_{1}, n_{3}$)
    \\ $edge(e_{3})$= ($n_{1}, n_{4}$)
 \\$lbl(n_{1}) =$ \{"Resource","Person"\}
    \\$\forall i \in$ \{$2,3,4$\}; $lbl(n_{i}) =$ \{"Resource"\}
  \\$\sigma(n_{1})$ = \{("url", "http://example.com/Monica"), \\("name", "Monica")\}
  \\$\sigma(n_{2})$ = \{("url", "http://www.Monicahompage.org")\}
    \\$\sigma(n_{3})$ = \{("url", "http://example.com/Management")\}
      \\$\sigma(n_{4})$ = \{("url", "http://example.com/Programming")\}
  
  \\ $\sigma(e_{1}) =$\{("type,\\ "$localName$(http://example.com/homepage)")\}
  \\ $\forall i \in$ \{$2,3$\}; $\sigma(e_{i}) =$ \{("type,\\"$localName($http://example.com/hasSkill)")\}
  
\end{tabular}
}
 & \multicolumn{1}{l|}{  \begin{tabular}[l]{@{}l@{}} Parsing error  \end{tabular} }
 
 & \multicolumn{1}{l|} {
 \begin{tabular}[l]{@{}l@{}}
 Not supported.
 \end{tabular} }
 
 \\ \hline 
 
  \multicolumn{1}{|p{1.5cm}|} { \centering Case 7} &
  \multicolumn{1}{l|}{ 
\begin{tabular}[l]{@{}l@{}} 

@prefix ex: <http://example.com/> .
\\
\\ ex:alice  a ex:Artist .
\\ex:alice  a ex:Author .		
\end{tabular}
 } 
 &
 \multicolumn{1}{l|}{ 
 \begin{tabular}[l]{@{}l@{}} 

 \\  $N=$\{$n_{1}$\}
  \\ $E= \emptyset$
 \\ $lbl_{n}(n_{1}) =$ \{"Resource", "Author",
 "Artist"\}
 \\ $\sigma(n_{1})$= \{("url","http://example.com/alice")\}
\end{tabular}
}
 & 
 
  \multicolumn{1}{l|} {\begin{tabular}[l]{@{}l@{}}  
$N= $\{$ n_{1}, n_{2}, n_{3}$\}
\\  $E=$\{$e_{1}, e_{2}$\}
 \\ $edge(e_{1})$= ($n_{1}, n_{2}$)
  \\ $edge(e_{2})$= ($n_{1}, n_{3}$)
 \\ $\forall i \in$ \{$1,2$\}; $lbl(e_{i})=$\{"http://www.w3.org/\\1999/02/22-rdf-syntax-ns\#type"\}
 \\ $\sigma(n_{1})$= \{("Kind","IRI"), \\("IRI","http://example.com/alice")\}
  \\ $\sigma(n_{2})$= \{("Kind","IRI"), \\("IRI", "http://example.com/Artist" )\}
   \\ $\sigma(n_{3})$= \{("Kind","IRI"),\\ ("IRI", "http://example.com/Author")\}
\end{tabular} }

 & 
     \multicolumn{1}{l|}{ 
\begin{tabular}[l]{@{}l@{}}  
  \\ $N=$\{$n_{1}, n_{2}, n_{3}$\}
\\  $E= $\{$e_{1}, e_{2}$\}
 \\ $edge(e_{1})$= ($n_{1}, n_{2}$)
  \\ $edge(e_{2})$= ($n_{1}, n_{3}$)
 \\ $\forall i \in$ \{$1,2$\}; $lbl(n_{i})=$ \{"Resource"\}
 \\ $\forall i \in$ \{$1,2$\}; $lbl(e_{i})=$ \{"ObjectProperty"\}
 
\\ $\sigma(n_{1})$= \{("iri","http://example.com/alice")\}
\\  $\sigma(n_{2})$= \{("iri","http://example.com/Artist")\}
 \\   $\sigma(n_{3})$= \{("iri","http://example.com/Author")\}
 \\ $\forall i \in$ \{$1,2$\}; $\sigma(e_{i})$= \{("type","http://www.w3.org\\/1999/02/22-rdf-syntax-ns\#type")\}
 
\end{tabular}
 }

 \\ \hline
 \multicolumn{5}{|c|} {\textbf{ RDF-star}}

\\ \hline
  \multicolumn{1}{|p{1.5cm}|} { \centering Case 8} & 
   
    \multicolumn{1}{l|}{ 
\begin{tabular}[l]{@{}l@{}} 
\\@prefix ex: <http://example.com/> .
\\
\\ <<ex:Mary ex:likes ex:Matt>>
\\ ex:certainty 0.5 .				
\end{tabular}
 }
 
   &  \begin{tabular}[l]{@{}l@{}}  
 \\   $N=$\{$n_{1}, n_{2}$\}
\\  $E= $\{$e_{1}$\}
 \\ $edge(e_{1})$= ($n_{1}, n_{2}$)
 \\ \\$\forall i \in$ \{$1,2$\}; $lbl(n_{i}) =$ \{"Resource"\}
 \\$\sigma(n_{1})$= \{("url","
"http://example.com/Mary")\}
 \\$\sigma(n_{2})$= \{("url","
"http://example.com/Matt")\}
 \\$\sigma(e_{1})$= \{("type","\\
 $localName$(http://example.com/likes)"),
 \\ ("certainty", 0.5)\}
\end{tabular}

& 

 \multicolumn{1}{l|} {\begin{tabular}[l]{@{}l@{}}  
$N= $\{$ n_{1}, n_{2}$\}
\\  $E=$\{$e_{1}$\}
 \\ $edge(e_{1})$= ($n_{1}, n_{2}$)
 \\$lbl_{e}(e_{1})=$\{ "http://example.com/likes"\}
 \\ $\sigma(n_{1})$=\{("Kind","IRI"),
\\ ("IRI", "http://example.com/Mary")\}
  \\ $\sigma(n_{2})$=\{("Kind","IRI"),\\
  ("IRI","http://example.com/Matt")\}
  \\ $\sigma(e_{1})$= \\\{("http://example.com/certainty:decimal", 0.5)\}

\end{tabular}
}
& 

\multicolumn{1}{l|} {\begin{tabular}[l]{@{}l@{}}
 Not supported.
 \end{tabular} }         \\ \hline
 
   \multicolumn{1}{|p{1.5cm}|} { \centering Case 9}  & 
  
   \multicolumn{1}{l|}{ 
\begin{tabular}[l]{@{}l@{}} 
\\@prefix ex: <http://example.com/> .
\\
\\ <<ex:Mark ex:age 28>> \\ 
    ex:certainty 1 .				
\end{tabular}
 }
  &\multicolumn{1}{l|}{ 
Triple is ignored.
 }
 &
   \multicolumn{1}{l|} {\begin{tabular}[l]{@{}l@{}}  
\\$N= $\{$ n_{1}, n_{2}$\}
\\  $E=$\{$e_{1}$\}
 \\ $edge(e_{1})$= ($n_{1}, n_{2}$)
 \\$lbl(e_{1})=$\{ "http://example.com/age"\}
 \\ $\sigma(n_{1})$= \{("Kind","IRI"), ("IRI", \\
http://example.com/Mark)\}
  \\ $\sigma(n_{2})$= \{("Kind","Literal"),
 \\ ("Literal", 28)\}
  \\ $\sigma(e_{1})$= \{("certainty", 1)\}

\end{tabular}
}

 & \multicolumn{1}{l|} {\begin{tabular}[l]{@{}l@{}}
 Not supported.
 \end{tabular} }           \\ \hline
 
  \multicolumn{1}{|p{1.5cm}|} { \centering Case 10} &

 \multicolumn{1}{l|}{ 
\begin{tabular}[l]{@{}l@{}} 
\\@prefix ex: <http://example.com/> .
\\
\\ ex:bobshomepage ex:source \\ << ex:mainPage ex:writer ex:alice >> .				
\end{tabular}
 }

 & 
  \multicolumn{1}{l|}{ 
 
Triple is ignored.
 
 }  & \multicolumn{1}{l|}{  Parsing error} & \multicolumn{1}{l|} {\begin{tabular}[l]{@{}l@{}}
 Not supported.
 \end{tabular} } \\ \hline
 
  \multicolumn{1}{|p{1.5cm}|} { \centering Case 11.1}& \multicolumn{1}{l|}{ 
\begin{tabular}[l]{@{}l@{}} 
\\@prefix ex: <http://example.com/> .
\\
\\ << ex:mainPage ex:writer ex:alice >> \\ ex:source  ex:bobshomepage .				
\end{tabular}
 }
 & 
 \multicolumn{1}{l|}{ 
Triple is ignored. }

 & 
    \multicolumn{1}{l|} {\begin{tabular}[l]{@{}l@{}}  
\\ $N= $\{$ n_{1}, n_{2}$\}
\\  $E=$\{$e_{1}$\}
 \\ $edge(e_{1})$= \{$n_{1}, n_{2}$\},
 \\$lbl(e_{1})=$ \{"http://example.com/writer"\}
 \\ $\sigma(n_{1})$= \{("Kind","IRI"), \\
 ("IRI","http://example.com/mainPage")\}
  \\ $\sigma(n_{2})$= \{("Kind","IRI"), \\
  ("IRI","http://example.com/alice")\}
  \\ $\sigma(e_{1})$= \{("http://example.com/source:String",\\"http://example.com/bobshomepage"\}
\end{tabular}
}

  & \multicolumn{1}{l|} {\begin{tabular}[l]{@{}l@{}}
 Not supported.
 \end{tabular} }     \\ \hline
 
 \multicolumn{1}{|p{1.5cm}|} { \centering Case 11.2} & 
 \multicolumn{1}{l|}{ 
\begin{tabular}[l]{@{}l@{}} 
\\@prefix ex: <http://example.com/> .
\\
\\ <<ex:alice ex:friend ex:bob>> \\ ex:mentionedBy ex:Alex.
\\ ex:Alex ex:age 25. 				
\end{tabular}
 }
 &
 
 \multicolumn{1}{l|}{ 
 \begin{tabular}[l]{@{}l@{}} 
 Nested Triple is ignored.
 \\   $N=$\{$n_{1}$\}
  \\$E= \emptyset$
 \\$lbl(n_{1}) =$ \{"Resource"\}
 \\$\sigma$($n_{1}$) = \{("url","http://example.com/Alex")\\("age", 25)\}

\end{tabular}
}

 &
    \multicolumn{1}{l|} {\begin{tabular}[l]{@{}l@{}}  
\\ $N= $\{$ n_{1}, n_{2}, n_{3}, n_{4}$\}
\\  $E=$\{$e_{1}, e_{2}$\}
 \\ $edge(e_{1})$= \{$n_{3}, n_{4}$\}
  \\ $edge(e_{2})$= \{$n_{1}, n_{2}$\}
 \\$lbl(e_{1})=$ \{"http://example.com/age"\}
 \\$lbl(e_{2})=$ \{"http://example.com/friend"\}
  \\ $\sigma(n_{1})$=\{("Kind","IRI"), \\
  ("IRI", "http://example.com/alice" )\}
  \\ $\sigma(n_{2})$=\{("Kind","IRI"),\\
  ("IRI", "http://example.com/bob")\}
  
  \\ $\sigma(n_{3})$=\{("Kind","IRI") ,\\ 
  ("IRI","http://example.com/Alex")\}
 \\ $\sigma(n_{4})$= \{("Kind","Literal"),
 \\("Literal", 25)\}
  \\ $\sigma(e_{2})$= \\ \{("http://example.com/mentionedBy:String", \\"http://example.com/Alex")\}

\end{tabular}
}

 & \multicolumn{1}{l|} {\begin{tabular}[l]{@{}l@{}}
Not supported.

 \end{tabular} }           \\ \hline

 \multicolumn{1}{|p{1.5cm}|} { \centering Case 12.1} &  \multicolumn{1}{l|}{ 
\begin{tabular}[l]{@{}l@{}} 
\\@prefix ex: <http://example.com/> .
\\@prefix rdf: <http://www.w3.org/\\1999/ 02/22-rdf-syntax-ns\#>.
\\
\\ << ex:mainPage ex:writer ex:alice >> \\ rdf:type  ex:bobshomepage .				
\end{tabular}
 }
 &
 \multicolumn{1}{l|}{ 
 
Triple is ignored.
 
 }  & 
 
    \multicolumn{1}{l|} {\begin{tabular}[l]{@{}l@{}}  
\\ $N= $\{$ n_{1}, n_{2}$\}
\\  $E=$\{$e_{1}$\}
 \\ $edge(e_{2})$= ($n_{1}, n_{2}$)
 \\$lbl(e_{1})=$\{"http://example.com/writer"\}
 \\ $\sigma(n_{1})$= \{("Kind","IRI"),\\
 ("IRI", " http://example.com/mainPage")\}
  \\ $\sigma(n_{2})$= \{("Kind","IRI"),
  \\ ("IRI","http://example.com/alice" )\}
  \\ $\sigma(e_{1})$= \{("http://www.w3.org/1999/02/\\22-rdf-syntax-ns\#type:String", \\(" http://example.com/bobshomepage")\}

\end{tabular}
}
  & \multicolumn{1}{l|} {\begin{tabular}[l]{@{}l@{}}
 Not supported.
 \end{tabular} }           
 \\ \hline
 \multicolumn{1}{|p{1.5cm}|} { \centering Case 12.2} & 
 
  \multicolumn{1}{l|}{ 
\begin{tabular}[l]{@{}l@{}} 
\\@prefix ex: <http://example.com/> .
\\@prefix rdf: <http://www.w3.org/\\1999/ 02/22-rdf-syntax-ns\#>.
\\
\\ <<ex:lara rdf:type ex:writer >> \\ ex:owner ex:Journal.				
\end{tabular}
 }
 
 & \multicolumn{1}{l|}{ 
 
Triple is ignored.
 
 }
 & 
 \multicolumn{1}{l|} {\begin{tabular}[l]{@{}l@{}}  
\\$N= $\{$ n_{1}, n_{2}$\}
\\  $E=$\{$e_{1}$\}
 \\ $edge(e_{2})=$ ($n_{1}, n_{2}$)
 \\$lbl(e_{1})=$ "http://www.w3.org/\\1999/02/22-rdf-syntax-ns\#type"
 \\ $\sigma(n_{1})$= \{("Kind","IRI"),\\ ("IRI","http://example.com/lara")\}
  \\ $\sigma(n_{2})$= \{("Kind","IRI"),\\ ("IRI","http://example.com/writer")\}
  \\ $\sigma(e_{1})$= \{("http://example.com/owner:String",\\ "http://example.com/Journal)"\}

\end{tabular}
}
  
 & \multicolumn{1}{l|} {\begin{tabular}[l]{@{}l@{}}
 Not supported.
 \end{tabular} }           \\ \hline

 \multicolumn{1}{|p{1.5cm}|} { \centering Case 13} & 
 
  \multicolumn{1}{l|}{ 
\begin{tabular}[l]{@{}l@{}} 
\\@prefix ex: <http://example.com/> .
\\
\\ <<<<ex:Steve ex:position "CEO">> \\ ex:mentionedBy ex:book>> \\ ex:source ex:journal.			
\end{tabular}
 }
 
 &\multicolumn{1}{l|}{ 
 
Triple is ignored.
 
 }  & \multicolumn{1}{l|}{ Parsing error} & \multicolumn{1}{l|} {\begin{tabular}[l]{@{}l@{}}
 Not supported.
 \end{tabular} }           \\ \hline
  \multicolumn{1}{|p{1.5cm}|} { \centering Case 14.1} &   \multicolumn{1}{l|}{ 
\begin{tabular}[l]{@{}l@{}} 
\\@prefix ex: <http://example.com/> .
\\
\\ex:college\_page ex:subject \\ \hspace{10mm} "Info\_Page"; 
        \\ ex:subject "aau\_page" .
\end{tabular}
 }
 &

 \multicolumn{1}{l|}{ 
 \begin{tabular}[l]{@{}l@{}} 

 \\  $N=$\{$n_{1}$\}
 \\$E= \emptyset$
 \\$lbl(n_{1}) =$ \{"Resource"\}
  \\$\sigma$($n_{1}$) = \{("url",\\"http://example.com/college\_page" ), ("subject", \\ \{"aau\_page", "Info\_Page"\})\}

\end{tabular}
}

&

 \multicolumn{1}{l|} {\begin{tabular}[l]{@{}l@{}}  
\\ $N= $\{$ n_{1}, n_{2}, n_{3}$\}
\\  $E=$\{$e_{1}, e_{2}$\}
  \\ $edge(e_{1})$= ($n_{1}, n_{2}$)
  \\ $edge(e_{2})$= ($n_{1}, n_{3}$)
 \\ $\forall i \in$ \{$1,2$\}; $lbl(e_{i})=$ \\ \{"http://example.com/subject"\}
 \\ $\sigma(n_{1})$= \{("Kind","IRI"), 
 ("IRI",\\ "http://example.com/college\_page")\}
  \\ $\sigma(n_{2})$= \{("Kind","Literal"),
 \\("Literal","Info\_Page")\}
   \\ $\sigma(n_{3})$= \{("Kind","Literal"),
 \\("Literal","aau\_page")\}

\end{tabular}
}
 &
    \multicolumn{1}{l|}{ 
\begin{tabular}[l]{@{}l@{}}  
\\   $N=$\{$n_{1}, n_{2}, n_{3}$\}
\\  $E= $\{$e_{1}, e_{2}$\}
 \\ $edge(e_{1})$= ($n_{1}, n_{2}$)
 \\ $edge(e_{2})$= ($n_{1}, n_{3}$)
 \\ $lbl(n_{1}) =$ \{"Resource"\}
\\  $\forall i \in$ \{$1,2$\}; $lbl(n_{i}) =$ \{"Literal"\}
 \\ $\forall i \in$ \{$1,2$\}; $lbl(e_{i}) =$ \{"DatatypeProperty"\}
 \\$\sigma(n_{n})$= \{("iri",\\ "http://example.com/college\_page)\}
 \\ $\forall i \in$ \{$1,2$\}; $\sigma(e_{i})$= \{("type", \\"http://example.com/subject")\}
   \\ $\sigma(n_{2})$= \{("value", "Info\_Page")\}
     \\ $\sigma(n_{3})$= \{("value", "aau\_page")\} 

\end{tabular}
 }
 
 \\ \hline

    \multicolumn{1}{|p{1.5cm}|} { \centering Case 14.2} &   \multicolumn{1}{l|}{ 
\begin{tabular}[l]{@{}l@{}} 
\\@prefix ex: <http://example.com/> .
\\
\\
<<ex:Mary ex:likes ex:Matt>> \\ \hspace{10mm}ex:certainty 0.5 .
\\<<ex:Mary ex:likes ex:Matt>> \\ \hspace{10mm}ex:certainty 1 .
\end{tabular}
 }
 &

 \multicolumn{1}{l|}{ 
 \begin{tabular}[l]{@{}l@{}} 

 \\  $N=$\{$n_{1}, n_{2}$\}
 \\  $E=$\{$e_{1}$\}
\\  $\forall i \in$ \{$1,2$\}; $lbl(n_{i}) =$ \{"Resource"\}

  \\$\sigma$($n_{1}$) = \{("url",\\"http://example.com/Mary" )\}
   \\$\sigma$($n_{2}$) = \{("url",\\"http://example.com/Matt" )\}
    \\$\sigma(e_{1})$= \{("type","\\
 $localName$(http://example.com/likes)"),
 \\ ("certainty", 1)\}
  
\end{tabular}
}

&

 \multicolumn{1}{l|} {\begin{tabular}[l]{@{}l@{}}  
\\ $N= $\{$ n_{1}, n_{2}$\}
\\  $E=$\{$e_{1}$\}
  \\ $edge(e_{1})$= ($n_{1}, n_{2}$)
 \\  $lbl(e_{1})=$ \\ \{"http://example.com/likes"\}
 \\ $\sigma(n_{1})$= \{("Kind","IRI"), \\
 ("IRI", "http://example.com/Mary")\}
  \\ $\sigma(n_{2})$= \{("Kind","IRI"),
 \\("IRI","http://example.com/Matt")\}
 
  \\ $\sigma(e_{1})$=\\ \{("http://example.com/certainty:decimal",1)
 \}
  
\end{tabular}
}
 &
\multicolumn{1}{l|} {\begin{tabular}[l]{@{}l@{}}
 Not supported.
 \end{tabular} }  
 
 \\ \hline

   \multicolumn{1}{|p{1.5cm}|} { \centering Case 15.1} &
  \multicolumn{1}{l|}{ 
\begin{tabular}[l]{@{}l@{}} 
\\ @prefix ex: <http://example.com/> .
\\
\\<<ex:Mary ex:likes ex:Matt>> \\ \hspace{10mm} ex:certainty 0.5 .
\\<<ex:Mary ex:likes ex:Matt>>\\ \hspace{10mm}ex:source "text" .

\end{tabular}
 } 
 &
 \multicolumn{1}{l|}{ 
 \begin{tabular}[l]{@{}l@{}} 

 \\  $N=$\{$n_{1}, n_{2}$\}
 \\  $E=$\{$e_{1}$\}
\\  $\forall i \in$ \{$1,2$\}; $lbl(n_{i}) =$ \{"Resource"\}

  \\$\sigma$($n_{1}$) = \{("url",\\"http://example.com/Mary" )\}
   \\$\sigma$($n_{2}$) = \{("url",\\"http://example.com/Matt" )\}
    \\$\sigma(e_{1})$= \{("type","\\
 $localName$(http://example.com/likes)"),
 \\ ("certainty", 1),
 ("source", "text")\}
  
\end{tabular}
}

 & 
 
 \multicolumn{1}{l|} {\begin{tabular}[l]{@{}l@{}}  
\\ $N= $\{$ n_{1}, n_{2}$\}
\\  $E=$\{$e_{1}$\}
  \\ $edge(e_{1})$= ($n_{1}, n_{2}$)
 \\  $lbl(e_{1})=$ \\ \{"http://example.com/likes"\}
 \\ $\sigma(n_{1})$= \{("Kind","IRI"), \\
 ("IRI", "http://example.com/Mary")\}
  \\ $\sigma(n_{2})$= \{("Kind","IRI"),
 \\("IRI","http://example.com/Matt")\}
 
  \\ $\sigma(e_{1})$=\\ \{("http://example.com/certainty:decimal",0.5),\\
  ("http://example.com/source:string","text")\}

\end{tabular}
}

 & 
\multicolumn{1}{l|} {\begin{tabular}[l]{@{}l@{}}
 Not supported.
 \end{tabular} }

 \\ \hline

  \multicolumn{1}{|p{1.5cm}|} { \centering Case 15.2} &
  \multicolumn{1}{l|}{ 
\begin{tabular}[l]{@{}l@{}} 
\\@prefix ex: <http://example.com/> .
\\
\\<<ex:Mary ex:likes ex:Matt>>\\
          \hspace{10mm}ex:certainty 0.5 .
 \\\hspace{2mm}ex:Mary  ex:likes ex:Matt .
	
\end{tabular}
 } 
 &
\multicolumn{1}{l|}{ 
 \begin{tabular}[l]{@{}l@{}} 

 \\  $N=$\{$n_{1}, n_{2}$\}
 \\  $E=$\{$e_{1}$\}
\\  $\forall i \in$ \{$1,2$\}; $lbl(n_{i}) =$ \{"Resource"\}

  \\$\sigma$($n_{1}$) = \{("url",\\"http://example.com/Mary" )\}
   \\$\sigma$($n_{2}$) = \{("url",\\"http://example.com/Matt" )\}
    \\$\sigma(e_{1})$= \{("type","\\
 $localName$(http://example.com/likes)"),
 \\ ("certainty", 0.5)\}
  
\end{tabular}
}

 & 
 
 \multicolumn{1}{l|} {\begin{tabular}[l]{@{}l@{}}  
\\ $N= $\{$ n_{1}, n_{2}$\}
\\  $E=$\{$e_{1}$\}
  \\ $edge(e_{1})$= ($n_{1}, n_{2}$)
 \\  $lbl(e_{1})=$ \\ \{"http://example.com/likes"\}
 \\ $\sigma(n_{1})$= \{("Kind","IRI"), \\
 ("IRI", "http://example.com/Mary")\}
  \\ $\sigma(n_{2})$= \{("Kind","IRI"),
 \\("IRI","http://example.com/Matt")\}
 
  \\ $\sigma(e_{1})$=\\ \{("http://example.com/certainty:decimal",0.5)\}

\end{tabular}
}
 & 
\multicolumn{1}{l|} {\begin{tabular}[l]{@{}l@{}}
 Not supported.
 \end{tabular} }         

 \\ \hline

\caption{This table presents the transformation results from RDF-star graphs into property graphs. 
The input for each case is an RDF set of triples (Triples Column). The output corresponds to the transformation by the three evaluated approaches: Neosemantics, RDF-Star Tools, and RDF2PG.}
\label{tab:mapping}
\end{longtable} 